\documentclass[11pt]{article}

\usepackage{amssymb}

\usepackage{graphicx}

\usepackage[english,francais]{babel}

\newfont{\ensmathquatorze}{msbm10 scaled 1400}
\newfont{\ensmathonze}{msbm10 scaled 1100}
\newfont{\ensmathdix}{msbm10}
\newfont{\ensmathneuf}{msbm10 scaled 833}
\newfont{\ensmathhuit}{msbm10 scaled 694}
\newfam\ensmathfam                        
\textfont\ensmathfam=\ensmathonze        
\scriptfont\ensmathfam=\ensmathdix       
\scriptscriptfont\ensmathfam=\ensmathhuit
\def\ensmf{\fam\ensmathfam\ensmathonze}         

\def\be{\begin{equation}}
\def\ee{\end{equation}}

\def\bea{\begin{eqnarray}}
\def\eea{\end{eqnarray}}
\def\beann{\begin{eqnarray*}}
\def\eeann{\end{eqnarray*}}

\def\eqdef{\stackrel{\mbox{\tiny def}}{=}}     
\newcommand{\ket}[1]{|\kern.3ex#1\kern.3ex\rangle}
\newcommand{\bra}[1]{\langle\kern.3ex #1 \kern.3ex|}
\newcommand{\APPROX}[1]{                
   {{\raisebox{-.3cm}{$\textstyle\simeq$}} \atop {\scriptstyle{#1}}}}

\newcommand{\EXP}[1]{{\mbox{\large e}}^{#1}}         

\newcommand{\re}{\mathop{\mathrm{Re}}\nolimits}      
\newcommand{\tr}[1]{\mathop{\mathrm{Tr}}\nolimits\left\{ #1 \right\}}  
\newcommand{\cotg}{\mathop{\mathrm{cotg}}\nolimits}  

\def\NN{{\ensmf N}}                 
\def\ZZ{{\ensmf Z}}                 
\def\RR{{\ensmf R}}                 

\def\I{{\rm i}}                  
\def\D{{\rm d}}                  
\def\Dc{{\rm D}}                 

\newcommand\ab{{\alpha\beta}}
\newcommand\ba{{\beta\alpha}}

\begin{document}

\selectlanguage{english}

\title{Local Friedel sum rule on graphs}

\author{Christophe Texier$^{(a)}$ and Markus B\"uttiker$^{(b)}$}

\date{\today}

\maketitle

\hspace{2cm}
\begin{minipage}[t]{12cm}
{\small
$^{(a)}$Laboratoire de Physique Th\'eorique et Mod\`eles Statistiques.

Universit\'e Paris-Sud,
B\^at. 100, F-91405 Orsay Cedex, France.

\hspace{3cm}and

Laboratoire de Physique des Solides.

Universit\'e Paris-Sud,
B\^at. 510, F-91405 Orsay Cedex, France.

\vspace{0.5cm}

$^{(b)}$D\'epartement de Physique Th\'eorique. Universit\'e de Gen\`eve.

24, quai Ernest Ansermet. CH-1211 Gen\`eve 4. Switzerland.
}
\end{minipage}

\begin{abstract}
We consider graphs made of one-dimensional wires connected at vertices and
on which may live a scalar potential. We are interested in a scattering
situation where the graph is connected to infinite leads.
We investigate relations between the scattering matrix and the continuous 
part of the local density of states, the injectivities, emissivities and 
partial local density of states.
Those latter quantities can be obtained by attaching an extra lead at the 
point of interest and by investigating the transport in the limit of zero 
transmission into the additional lead.
In addition to the continuous part related to the scattering states,
the spectrum of graphs may present a discrete part related to states
that remain uncoupled to the external leads.
The theory is illustrated with the help of a few simple examples.

\end{abstract}

\noindent
PACS~: 03.65.Nk, 73.23.-b





\section{Introduction\label{sec:Intro}}

Thanks to the powerful experimental techniques used in mesoscopic physics
during the past 20 years, many interesting and fundamental effects
have been investigated on systems that can be modelized by graphs~:
networks of wires through which an electrical current can flow.
The most famous example is of course the Aharonov-Bohm oscillations in  a
coherent metallic ring \cite{ButImrLan83,GefImrAzb84,ButImrAzb84,But85}
(see also the
excellent review \cite{WasWeb86}), but there have also been many other
realizations of graphs like the recent ones devoted to the
Aharonov-Bohm cage effect
\cite{VidMonDou00,PanAbiSerFouButVid01,NauFaiMaiEti01,Nau01}.
Graphs which could seem at first sight to be oversimplified models for
mesoscopic networks, succeeded in those cases to describe the interesting
physical effects, which explain why they are so widely used in many
theoretical works.

Among all the useful concepts of mesoscopic physics the scattering
approach
plays a central role. It provides a powerful tool to study many physical
quantities related to transport, noise, etc.
A very important concept of the scattering theory is the one related to
the Krein-Friedel relation, or Friedel sum rule \cite{Fri52,Kre53},
which establishes a relation between the scattering and the spectral
properties\footnote{
  In fact the idea of relating the spectral properties to the scattering
  properties goes back to 1937 when Beth and Uhlenbeck \cite{BetUhl37}
  related the second virial coefficient of a gas of interacting particles
  to the phase shifts of the 2-body scattering problem (see also
  \cite{Hua63} or \S77 of \cite{LanLif66e}).
}.
The use of this kind of relations in mesoscopic physics allows to express,
for instance, the charge distribution in terms of the scattering
properties. However
it permits to consider only the total charge in the scattering region
in an equilibrium situation.
Several works have been devoted to build corresponding concepts
to describe out-of-equilibrium situations and also to relate not only
global but {\it local} quantities to the scattering (like the local density
of states). This has lead to the introduction of the concepts of partial
local density
of states, injectivities, emissivities, etc
\cite{But93,GasChrBut96,GraBut97}.
Let us also mention the recent work on the relation between scattering
and local density of states for the particular case of
quasi-one-dimensional systems \cite{SouSuz02,SchTitBroBee02}.

The scattering theory on graphs has attracted the interest of many authors
among which we can quote
\cite{GerPav88,AvrSad91,Ada92,KosSch99,KotSmi00,TexMon01,Tex02}.
The purpose of this article is to discuss the relation between scattering
properties of the graph and the local quantities mentioned above.


\section{Motivation}

The local density of states (LDoS) is by definition~:
$\rho(x;E)=\bra{x}\delta(E-H)\ket{x}$.
If we consider a graph ${\cal G}$ connected to infinite leads, its
spectrum
is continuous. We can define the DoS of the scattering region (the graph)
as
$\rho_{\cal G}(E)=\int_{\cal G}\D x\,\rho(x;E)$, where the integral runs
over
all the bonds of the graph (by convention, we do not include the infinite
leads in what we call ``graph ${\cal G}$''). It is well known that
this object is related to the scattering properties through the
Krein-Friedel
relation (or Friedel sum rule). Graphs have the particularity (which
does not occur in 1d for example) that due to certain symmetries some
states may remain uncoupled to the external leads.
The presence of such localized states leads to the existence of a
discrete
part in the spectrum, superimposed on the continuous part~:
$\rho_{\cal G}(E)=\rho_{\rm reg}(E)+\rho_{\rm dis}(E)$ where
$\rho_{\rm reg}(E)$ is the contribution of the stationary scattering
states
and $\rho_{\rm dis}(E)=\sum_n g_n\delta(E-E_n)$ is the contribution of the
localized states (see appendix \ref{app:spectrum}).
The existence of these two contributions was already
noticed by J.-P.~Roth in a work \cite{Rot84} in which he extended
the trace formula obtained for closed graphs without potential
\cite{Rot83}
to open graphs connected to semi infinite wires.
What is rather unusual is that the continuous part and the discrete part
live on the same intervals of energy. If the wave function is chosen
continuous
at the vertices, the
absence of hybridization of the localized states with the states of the
continuum occurs when some states vanish at all the vertices
to which external leads are connected.
It was shown in \cite{Tex02} that the Friedel sum rule has to be modified
since, obviously, only the continuous part is related to the scattering
matrix~:
$
\rho_{\rm reg}(E)=
\frac{1}{2\I\pi}
\left(
  \frac{\D}{\D E}\ln\det\Sigma+\frac1{4E}\tr{\Sigma-\Sigma^\dagger}
\right)
$
where $\Sigma$ is the scattering matrix of the graph.

Similarly, the LDoS can be separated into a continuous part related to the
stationary scattering states and a discrete part given by the localized
states (see appendix \ref{app:spectrum})~:
\be
\rho(x;E) = \sum_\alpha |\tilde\psi_E^{(\alpha)}(x)|^2
+ \sum_n \sum_{j=1}^{g_n} \delta(E-E_n)\,|\varphi_{n,j}(x)|^2
\ee
(if $x\notin{\cal G}$, only the first term remains since the localized
wave functions vanish outside the graph of course).
$\tilde\psi_E^{(\alpha)}(x)$ is the stationary scattering state
corresponding
to the injection of a plane wave at lead $\alpha$. The sum over $\alpha$
runs
over the $L$ vertices connected to leads.
The normalization is chosen to associate to those states a measure $\D E$.
The wave function $\varphi_{n,j}(x)$ is normalized to unity
in the graph, $j$ being a degeneracy label. $g_n$ is the number of the
localized states with energy $E_n$.

The modified Friedel sum rule is an example of a relation between a
``global'' quantity characterizing the graph (the regular part
$\rho_{\rm reg}(E)$ of the DoS $\rho_{\cal G}(E)$) and the scattering
matrix.
The purpose of the present work is to demonstrate some relations between
local quantities such as the local density of states (LDoS),
injectivities,
emissivities and partial LDoS, and the objects of scattering theory
(stationary states, scattering matrix) in the context of graphs.

In particular we will show that the first term of the LDoS is related to a
functional derivative of the scattering matrix. The second term of the
LDoS, if it exists, cannot be probed from the scattering properties.

\section{The scattering matrix}

We consider a graph ${\cal G}$ being the domain of the Schr\"odinger
operator
$H=-\Dc_x^2+V(x)$, where $\Dc_x=\D_x-\I A(x)$ is the covariant derivative.
The graph is a network of $B$ bonds joining $V$ vertices (denoted by greek
letters $\alpha,\beta,\ldots$). Each bond $(\ab)$ is identified with the
interval $[0,l_\ab]$ of $\RR$ so that a scalar function $\psi(x)$ defined
on the graph is characterized by its $B$ components $\psi_{(\ab)}(x)$.
The Schr\"odinger operator acts on functions $\psi(x)$ which are chosen,
in a first step, to be continuous at the vertices of the graph~:
$\psi_{(\ab)}(x=0)=\psi_\alpha$ for all vertices $\beta$ neighbours of
$\alpha$. We denote by $\psi_\alpha$ the value of the wave function at the
vertex. Additionally we must add a constraint on the derivatives of $\psi$
to ensure current conservation. As soon as continuity of $\psi(x)$
is required, the most
general additional condition is
$\sum_\beta a_\ab \Dc_x\psi_{(\ab)}(\alpha)=\lambda_\alpha\psi_\alpha$
where the presence of the connectivity matrix $a_\ab$ in the sum ensures
that it runs over the neighbouring vertices of $\alpha$.
The connectivity (or adjacency) matrix describes the topology of the
graph~:
$a_\ab=1$ if $\alpha$ and $\beta$ are connected by a bond, and $0$
otherwise.
The notation $\psi_{(\ab)}(\alpha)\equiv\psi_{(\ab)}(x=0)$ designates
the value of the component at the vertex.
$\lambda_\alpha$ is a real parameter that affects the scattering at the
vertex. It allows to interpolate between Neumann boundary conditions
($\lambda_\alpha=0$) and Dirichlet boundary
conditions ($\lambda_\alpha=\infty$, which
imposes $\psi_\alpha=0$). We can develop an intuition of the role of this
parameter by noting that if the vertex has coordination $2$, the boundary
condition describes a potential $\lambda_\alpha\delta(x)$ at the vertex.
Note that the transmission amplitude through the vertex is 
$2/(m_\alpha+\I\lambda_\alpha/k)$ where $m_\alpha$ is its coordination~;
the transmission is maximized for $\lambda_\alpha=0$.

Among the $V$ vertices of the graph, $L$ are connected to infinite leads.
The couplings to the leads can be chosen arbitrary such that we can go
continuously from a situation where the graph is coupled to the leads to a
situation where it becomes decoupled from some leads. This procedure
introduces
some discontinuity between the wave function at the extremity of the lead
and at the vertex of the graph to which the lead is connected
\cite{TexMon01}.
The scattering matrix $\Sigma$ is an $L\times L$ matrix that depends on the
energy $E=k^2$. It can be constructed by manipulating matrices that encode
the information about the graph (topology, potentials on the bonds,
lengths
of the bonds, magnetic fluxes, couplings to the leads) \cite{TexMon01}~:
\be\label{RES2}
\Sigma = -1 + 2 \, W \left( M + W^{\rm T}W\right)^{-1} W^{\rm T}
\:.\ee
The rectangular matrix $W$ encodes the information on the way the
graph is connected to leads~:
\be\label{RES4}
W_\ab=w_\alpha\,\delta_\ab
\ee
with $\alpha\in{\cal V}_{\rm ext}$ and $\beta\in{\cal V}$, where
${\cal V}=\{1,\cdots,V\}$ is the set of vertices and
${\cal V}_{\rm ext}$ the set of vertices connected to leads
(${\rm Card}({\cal V}_{\rm ext})=L$).
The parameter $w_\alpha\in\RR$ describes the coupling between the graph
and the lead at vertex $\alpha$. Its precise definition is given in
\cite{TexMon01}~: the transmission amplitude between the lead and
the graph is $2w_\alpha/(1+w_\alpha^2)$.

We call $x_\ab\in[0,l_\ab]$ the coordinate on the bond $(\ab)$, starting
from $\alpha$ (note that $x_\ab+x_\ba=l_\ab$, the length of the bond).
To describe the potentials on the bonds, we introduce two real functions
$f_\ab(x_\ab)$, $f_\ba(x_\ab)$~: the two linearly independent solutions of
the
Schr\"odinger equation $[E+\D_x^2-V_{(\ab)}(x)]f(x)=0$ on the bond,
satisfying boundary conditions~:
$f_\ab(0)=1$, $f_\ab(l_\ab)=0$,
$f_\ba(0)=0$ and $f_\ba(l_\ab)=1$.
For example, in the free case ($V(x)=0$), we have
$f_\ab(x_\ab)=\frac{\sin k(l_\ab-x_\ab)}{\sin kl_\ab}$ and
$f_\ba(x_\ab)=\frac{\sin kx_\ab}{\sin kl_\ab}$.

The matrix $M$ that contains all the information on the
isolated graph (potential on the bond, topology) is~:
\be\label{MJean}
M_\ab(-E) = \frac{\I}{\sqrt{E}}
\left(
  \delta_\ab
  \left[
    \lambda_\alpha -
    \sum_\mu a_{\alpha\mu} \frac{\D f_{\alpha\mu}}{\D
x_{\alpha\mu}}(\alpha)
  \right]
  + a_\ab \frac{\D f_\ab}{\D x_\ab}(\beta)\,\EXP{\I\theta_\ab}
\right)
\:,\ee
we introduced the obvious notation $f_\ab(0)\equiv f_\ab(\alpha)$,
$f_\ab(l_\ab)\equiv f_\ab(\beta)$, etc. $\theta_\ab$ is the magnetic flux
along the bond.
This matrix was introduced in the study of the spectral determinant of
isolated graphs \cite{AkkComDesMonTex00,Des00,Des01}.
Instead of encoding the information about the potential through the
functions
$f_\ab(x)$, it can be more conveniently related to the reflection and
transmission coefficients of each bond \cite{TexMon01}~:
\bea\label{RES3}
M_\ab &=&
\delta_\ab\left(\I\frac{\lambda_\alpha}{k}
+ \sum_\mu a_{\alpha\mu}
\frac{(1-r_{\alpha\mu})(1+r_{\mu\alpha})+t_{\alpha\mu}\,t_{\mu\alpha}}
     {(1+r_{\alpha\mu})(1+r_{\mu\alpha})-t_{\alpha\mu}\,t_{\mu\alpha}}
\right)
\nonumber \\ && \hspace{1cm}
- a_\ab\frac{2\,t_\ab}{(1+r_\ab)(1+r_\ba)-t_\ab\,t_\ba}
\:.\eea
These equations generalize the result known in the absence of the
potential \cite{AvrSad91}. In this latter case we recover from
(\ref{RES3}) the well-known matrix:
\be\label{fmM}
M_\ab = \I\,\delta_\ab\sum_\mu a_{\alpha\mu} \cotg kl_{\alpha\mu}
-a_\ab\frac{\I\,\EXP{\I\theta_\ab}}{\sin kl_\ab}
\:.\ee
We are now ready to discuss the extraction of
local information from the scattering matrix.


\section{Functional derivative of the scattering matrix}

The scattering matrix is a functional of the potential $V(x)$ and we are
now going to compute $\frac{\delta\,\Sigma}{\delta V(x)}$.
As a starting point it is useful to note that if a $\delta$ potential at
$x=x_0$ is added to the potential, the first perturbative correction to
the
scattering matrix is exactly the functional derivative~\cite{GasChrBut96}:
\be
\Sigma^\lambda \eqdef \Sigma[V(x)+\lambda\,\delta(x-x_0)]
= \Sigma[V(x)] + \lambda \frac{\delta\,\Sigma}{\delta V(x_0)}[V(x)]+\cdots
\ee
The advantage with graphs is that the addition of a $\delta$ potential
at $x$ is easily implemented~:
it is done by adding a vertex of weight $\lambda$ at $x$, obtaining a
graph
${\cal G}^\lambda$. The ``weight'' is the real parameter involved in the
mixed boundary conditions introduced above
\cite{AvrSad91,Avr95,AkkComDesMonTex00}.
As a consequence, the construction of the matrix $\Sigma^\lambda$ of
${\cal G}^\lambda$ in the vertex approach requires to consider
matrices $M$ and $W$ of size $(V+1)\times(V+1)$ and $L\times(V+1)$,
respectively. We call $W_x$ the matrix describing the coupling of the
graph to the leads when the additional vertex is at $x$ (the matrix $W_x$
has only an additional column of $0$'s compare to $W$) and $M^\lambda$ the
new
matrix $M$.
We have~:
\be
\Sigma^\lambda = -1 + 2\, W_x \frac{1}{M^\lambda+W_x^{\rm T}W_x}W_x^{\rm T}
\ee
From (\ref{MJean},\ref{RES3}) we see that the matrix $M^\lambda$ depends
linearly on $\lambda$~:
\be
M^\lambda = M_x + \frac{\I\lambda}{k}K_x
\ee
where $M_x$ is the matrix of the graph ${\cal G}^{\lambda=0}$. This graph
differs from ${\cal G}$ only by its number of vertices~: it possesses an
additional vertex of weight $\lambda=0$ at $x$. Since no scattering occurs
at this vertex when $\lambda=0$, we do not change the properties of the
graph, but only the size of the matrices describing it.
In the following we adopt notations such that all matrices with label $x$
refer to the graph ${\cal G}_x\eqdef{\cal G}^{\lambda=0}$.
The matrix $K_x$ contains only one non zero element coupling the vertex
$x$ to itself~:
\be
({K_x})_{\alpha\beta} = \delta_{\alpha\beta}\delta_{\alpha x}
\ee
where the indices run over the $V$ vertices of the initial graph and the
additional vertex at $x$.
Expanding the scattering matrix $\Sigma^\lambda$ in powers of $\lambda$ we
get~:
\be
\Sigma^\lambda = -1 + 2\, W_x \frac{1}{M_x+W_x^{\rm T}W_x}W_x^{\rm T}
- 2\,W_x \frac{1}{M_x+W_x^{\rm T}W_x}
\frac{\I\lambda}{k}K_x\frac{1}{M_x+W_x^{\rm T}W_x}W_x^{\rm T}
+\cdots
\ee
The first term is the scattering matrix $\Sigma = \Sigma^{\lambda=0}$
\be\label{sm}
\Sigma
= -1 + 2\, W_x \frac{1}{M_x+W_x^{\rm T}W_x}W_x^{\rm T}
= -1 + 2\, W \frac{1}{M+W^{\rm T}W}W^{\rm T}
\ee
and the second gives the functional derivative.
Then~:
\be\label{fd1}
\frac{\delta\,\Sigma}{\delta V(x)} =
-\frac{2\I}{k}W_x \frac{1}{M_x+W_x^{\rm T}W_x}
K_x\frac{1}{M_x+W_x^{\rm T}W_x}W_x^{\rm T}
\ee
This expression allows to compute explicitly the functional derivative
by manipulating matrices.


\section{Functional derivative and its relation to wave functions}

We now show the relation between the functional derivative of the
scattering
matrix and the stationary scattering state wave functions.
The wave function at the vertices of  ${\cal G}_x$ is a solution of
\cite{TexMon01}
\bea
1 + \Sigma               &=& W_x \Psi_x \\
W_x^{\rm T} (1 - \Sigma) &=& M_x \Psi_x
\:,\eea
where $\Psi_x$ is the $(V+1)\times L$ matrix gathering the wave function
$\psi_\mu^{(\alpha)}$ at vertex $\mu$ for the stationary scattering state
$\psi^{(\alpha)}(x)$~: the matrix element is by definition
$\Psi_{\mu\alpha}\equiv\psi^{(\alpha)}_\mu$.
The index $\mu$ runs over the $V$ vertices and the
additional vertex $x$. $\alpha$ runs over the $L$ vertices connected to
leads.
The wave function with the correct normalization is~:
\be
\tilde\psi^{(\alpha)}(x)=\frac{1}{\sqrt{4\pi k}}\,\psi^{(\alpha)}(x)
\:.\ee
In the previous papers \cite{TexMon01,Tex02} we have used the notation
$\tilde\psi^{(\alpha)}_E(x)$.
Here we omit the energy label $E$ to lighten the expressions.
It follows from the above equations that the wave function at the $V$
vertices
for the $L$ stationary scattering states is encoded in the matrix~:
\be
\Psi_x = 2 \frac{1}{M_x+W_x^{\rm T}W_x}W_x^{\rm T}
\ee
For the state associated to a measure $\D E$~:
\be\label{wf}
\tilde\Psi_x = \frac{1}{\sqrt{\pi k}}\,\frac{1}{M_x+W_x^{\rm
T}W_x}W_x^{\rm T}
\:.\ee
These matrix expressions are useful if we want to establish a relation
between
the functional derivative of the scattering matrix and wave functions.

\subsection{Time reversed graph}

If we consider a graph described by a matrix $M(\{\theta_{\alpha\beta}\})$
depending on magnetic fluxes $\theta_{\alpha\beta}$, it follows from the
construction of $M$
(see formula (48) and appendix A of \cite{TexMon01}) that the time
reversed graph (with all fluxes reversed) is described by the matrix
$M(\{-\theta_{\alpha\beta}\})=M(\{\theta_{\alpha\beta}\})^{\rm T}$.
The wave function at the vertices for the time-reversed graph
is then~:
\be\label{trwf}
\tilde\Psi^{t.r.}
= \frac{1}{\sqrt{\pi k}}\,\frac{1}{M^{\rm T}+W^{\rm T}W}W^{\rm T}
\:.\ee

\noindent
{\it Important remark}~: 
$\tilde\Psi^{t.r.}$ gives the wave function at the vertices for the 
time-reversed graph, which is related to the original graph
by reversing all the fluxes. 
It should not be confused with the wave function describing the time-reversed
motion of the electron (see also appendix \ref{app:TRS}).

\subsection{A first result}

We can now relate the functional derivative (\ref{fd1}) to the wave
function.
From the above remark we get~:
\be\label{fd2}
\frac{\delta\,\Sigma}{\delta V(x)} =
-2\I\pi (\tilde\Psi_x^{t.r.})^{\rm T} K_x\tilde\Psi_x
\ee
The matrix $K_x$ select the line in $\tilde\Psi_x$ associated with $x$ and
the corresponding column in $(\tilde\Psi_x^{t.r.})^{\rm T}$. Then
the matrix elements read~:
\be\label{fd3}
\frac{\delta\,\Sigma_{\alpha\beta}}{\delta V(x)} =
-2\I\pi \, \tilde\psi_x^{t.r.(\alpha)} \, \tilde\psi_x^{(\beta)}
\:.\ee
Since $\tilde\psi_x^{(\alpha)}$ is the wave function at vertex $x$ of
the graph, we could write more elegantly~:
\be\label{fd4}
\frac{\delta\,\Sigma_{\alpha\beta}}{\delta V(x)} =
-2\I\pi\, \tilde\psi^{t.r.(\alpha)}(x)\, \tilde\psi^{(\beta)}(x)
\:.\ee
For the scattering matrix element with indices interchanged we have
obviously
\be
\frac{\delta\,\Sigma_{\beta\alpha}}{\delta V(x)} =
-2\I\pi\, \tilde\psi^{t.r.(\beta)}(x)\, \tilde\psi^{(\alpha)}(x)
\:\ee
which shows that
\be
\frac{\delta\,\Sigma_{\alpha\beta}^{t.r.}}{\delta V(x)} =
\frac{\delta\,\Sigma_{\beta\alpha}}{\delta V(x)}
\:\ee
as required by the symmetry of the scattering matrix under
flux reversal.


\subsection{A compact formulation of the functional derivative}

All above formulae (\ref{fd1},\ref{fd2}) expressing the functional
derivative of the scattering
matrix involve matrices describing the graph ${\cal G}_x$. The purpose of
this paragraph is to simplify (\ref{fd1}) and express the functional
derivative in terms of matrices of smaller size, related to the original
graph ${\cal G}$.

We suppose that $x$ belongs to the
bond $(\ab)$ and we choose to organize the basis of $V+1$ vertices of
${\cal G}_x$ as $\{\cdots,\alpha,\beta\,|x\}$ to help the discussion.
The matrix of interest has the following structure~:
\be
M_x+W_x^{\rm T}W_x = \left(\begin{array}{cccc|c}
 & & & & \vdots \\
 & & & & 0      \\
 & & (M_x)_{\alpha\alpha}+w_\alpha^2 & 0 & (M_x)_{\alpha x} \\
 & & 0 & (M_x)_{\beta\beta}+w_\beta^2    & (M_x)_{\beta x}  \\ \hline
\cdots & 0 & (M_x)_{x\alpha} & (M_x)_{x\beta} & (M_x)_{xx}
\end{array}\right)
=\left(\begin{array}{c|c} A&B\\ \hline C&(M_x)_{xx}\end{array}\right)
\ee
Since the vertex $x$ is the only neighbour of $\alpha$ and $\beta$,
the part which is not written in the block $A$ is precisely the same as
in $M+W^{\rm T}W$. We have separated by a line the matrices into blocks
related to the vertices of ${\cal G}$ and the additional vertex $x$.
The matrices $W_x$ and $K_x$ read~:
\be
W_x = \left(\begin{array}{c|c}W&0\end{array}\right)
\hspace{2cm}
K_x = \left(\begin{array}{c|c}0&0\\\hline0&1\end{array}\right)
\:.\ee
To express the inverse of the matrix $M_x+W_x^{\rm T}W_x$, we note that
$A-B(M_x)_{xx}^{-1}C=M+W^{\rm T}W$. Then~:
\be
\left(M_x+W_x^{\rm T}W_x\right)^{-1}=
\left(\begin{array}{c|c}
(M+W^{\rm T}W)^{-1} & - (M+W^{\rm T}W)^{-1} B (M_x)_{xx}^{-1} \\ \hline
- (M_x)_{xx}^{-1} C (M+W^{\rm T}W)^{-1} & \cdots
\end{array}\right)
\:.\ee
After a little bit of algebra, we obtain from (\ref{fd1})~:
\be\label{fd5}
\frac{\delta\,\Sigma }{\delta V(x)}
=-\frac{2\I}{k}W\frac{1}{M+W^{\rm T}W} K(x)K(x)^\dagger
                    \frac{1}{M+W^{\rm T}W}W^{\rm T}
\ee
where the $V\times V$ matrix $K(x)K(x)^\dagger$ couples only the vertices
$\alpha$ and $\beta$~:
\be
K(x)K(x)^\dagger = \frac{1}{(M_x)_{xx}^2}
\left(\begin{array}{c}
\vdots \\
0 \\
(M_x)_{\alpha x}\\
(M_x)_{\beta x}
\end{array}\right)
\left(\begin{array}{cccc}
\cdots & 0 & (M_x)_{x\alpha}  & (M_x)_{x\beta}
\end{array}\right)
\:.\ee

We can also derive a more transparent  expression of the matrix
$K(x)K(x)^\dagger$
in terms of the two functions $f_\ab(x)$ and $f_\ba(x)$ introduced
above.
On the bond $(\ab)$ the wave function is
\be
\tilde\psi_{(\ab)}(x)=
  \tilde\psi_\alpha \EXP{\I\theta_\ab x/l_\ab} f_\ab(x)
+ \tilde\psi_\beta \EXP{-\I\theta_\ab (1-x/l_\ab)} f_\ba(x)
\:,\ee
where $x\equiv x_\ab\in[0,l_\ab]$ measures the distance from the vertex
$\alpha$.
The wave function of the time-reversed graph is obtained by changing the
sign of flux (the wave function at vertices is of course also affected by
this operation).
Then we can write
$\tilde\psi^{(\mu)}_{(\ab)}(x_\ab) = (K(x)^\dagger\tilde\Psi)_\mu$
where
\be
K(x)^\dagger=\left(\begin{array}{cccc}
  \cdots&0&f_\ab(x_\ab)\EXP{\I\theta_{\alpha x}}&
  f_\ba(x_\ab)\EXP{\I\theta_{\beta x}}
\end{array}\right)
\:.\ee
We have also~:
$\tilde\psi^{t.r.(\mu)}_{(\ab)}(x_\ab)
=((\tilde\Psi^{t.r.})^{\rm T}K(x))_\mu$.

If we now write
\be
-\frac{1}{2\I\pi}\frac{\delta\,\Sigma }{\delta V(x)}
= (\tilde\Psi^{t.r.})^{\rm T} K(x)K(x)^\dagger\tilde\Psi
\ee
We find from (\ref{fd4}) the new expression of the matrix involved in
(\ref{fd5})~:
\bea
K(x)K(x)^\dagger
&=&
\left(\begin{array}{c}
  \vdots\\0\\f_\ab(x_\ab)\EXP{-\I\theta_{\alpha x}}\\
  f_\ba(x_\ab)\EXP{-\I\theta_{\beta x}}
\end{array}\right)
\left(\begin{array}{cccc}
  \cdots&0&f_\ab(x_\ab)\EXP{\I\theta_{\alpha x}}&
  f_\ba(x_\ab)\EXP{\I\theta_{\beta x}}
\end{array}\right)
\:.\eea
We can extract some nontrivial relations between the elements of
$M_x$ and the functions $f_\ab(x_\ab)$ and $f_\ba(x_\ab)$~:
\be
f_\ab(x_\ab) = -\frac{(M_x)_{x\alpha}}{(M_x)_{xx}}
=\frac{ \frac{\D f_{x\alpha}}{\D x_{x\alpha}}(\alpha) }
      { \frac{\D f_{x\alpha}}{\D x_{x\alpha}}(x)
       +\frac{\D f_{x\beta}}{\D x_{x\beta}}(x) }
\:,\ee
which could have been demonstrated more directly by constructing
$f_\ab(x_\ab)$ in terms of $f_{\alpha x}(x_{\alpha x})$,
$f_{x\alpha}(x_{\alpha x})$ and $f_{x\beta}(x_{x\beta})$.


\section{Partial LDoS, injectivies, emissivities and LDoS}

\noindent{\bf Partial LDoS.}
The partial LDoS is defined as \cite{But93,GasChrBut96}
\be
\rho(\alpha,x,\beta) \eqdef - \frac1{4\I\pi}
\left(\Sigma^*_{\alpha\beta}
\frac{\delta\,\Sigma_{\alpha\beta}}{\delta V(x)}
-\frac{\delta\,\Sigma^*_{\alpha\beta}}{\delta V(x)}
\Sigma_{\alpha\beta}\right)
\:.\ee
It follows from (\ref{fd4}) that~:
\be
\rho(\alpha,x,\beta) =
\re\left[\Sigma_{\alpha\beta}^*\,
    \tilde\psi^{t.r.(\alpha)}(x)\, \tilde\psi^{(\beta)}(x)\right]
\:.\ee

\noindent{\bf Injectivies and emissivities.}
The injectivities are defined as
\be
\rho(x,\beta) \eqdef \sum_\alpha\rho(\alpha,x,\beta)
\ee
and the emissivities as
\be
\rho(\alpha,x) \eqdef \sum_\beta\rho(\alpha,x,\beta)
\:.\ee
They can be rewritten as~:
\be\label{inj1}
\rho(x,\alpha) = -\frac{1}{2\I\pi}
\left(\Sigma^\dagger\frac{\delta\,\Sigma}{\delta V(x)}
     \right)_{\alpha\alpha}
\ee
and
\be
\rho(\alpha,x) = -\frac{1}{2\I\pi}
\left(\frac{\delta\,\Sigma}{\delta V(x)}
      \Sigma^\dagger\right)_{\alpha\alpha}
\:.\ee
These two quantities are real thanks to the relations
$\frac{\delta\,\Sigma^\dagger}{\delta V(x)}\Sigma+
\Sigma^\dagger\frac{\delta\,\Sigma}{\delta V(x)}=0$
and
$\frac{\delta\,\Sigma}{\delta V(x)}\Sigma^\dagger+
\Sigma\frac{\delta\,\Sigma^\dagger}{\delta V(x)}=0$
coming from the unitariry~:
$\Sigma^\dagger\Sigma=\Sigma\Sigma^\dagger=1$.

We can now compute~:
\be
\Sigma^\dagger\frac{\delta\,\Sigma}{\delta V(x)} =
 -2\I\pi\, \Sigma^\dagger(\tilde\Psi_x^{t.r.})^{\rm T} K_x\tilde\Psi_x
=-2\I\pi\, \tilde\Psi_x^\dagger K_x\tilde\Psi_x
\:,\ee
where we have used (\ref{ur1}). Since
\be
(\tilde\Psi_x^\dagger K_x\tilde\Psi_x)_{\alpha\beta}
=\sum_{\mu,\nu}\tilde\psi_\mu^{(\alpha)\,*} \delta_{\mu\nu}\delta_{\mu x}
\tilde\psi_\nu^{(\beta)} = \tilde\psi^{(\alpha)\,*}_x\,
\tilde\psi^{(\beta)}_x
\ee
it follows that~:
\be\label{important}
-\frac{1}{2\I\pi}
\left(
  \Sigma^\dagger\frac{\delta\,\Sigma}{\delta V(x)}
\right)_{\alpha\beta}
=\tilde\psi^{(\alpha)}(x)^*\,\tilde\psi^{(\beta)}(x)
\ee
a relation that has been demonstrated in a different context and by
different means in Ref. \cite{But00}.
From (\ref{inj1}) we finally get for the injectivities~:
\be\label{inj2}
\rho(x,\alpha) = |\tilde\psi^{(\alpha)}(x)|^2
\:.\ee
The physical meaning of the injectivities is now clear~: it is the
contribution of the stationary scattering state incoming from $\alpha$ to
the continuous part of the LDoS.

We proceed similarly as above~:
\be
\frac{\delta\,\Sigma}{\delta V(x)}\Sigma^\dagger =
 -2\I\pi\, (\tilde\Psi_x^{t.r.})^{\rm T} K_x\tilde\Psi_x \Sigma^\dagger
=-2\I\pi\, (\tilde\Psi_x^{t.r.})^{\rm T} K_x (\tilde\Psi_x^{t.r.})^*
\:.\ee
Then
\be\label{emi}
\rho(\alpha,x) =
\left(\tilde\Psi_x^{t.r.}{}^{\rm T} K_x
      \tilde\Psi_x^{t.r.}{}^*\right)_{\alpha\alpha}
\ee
and we finally get
\be\label{wmi2}
\rho(\alpha,x) = |\tilde\psi^{t.r.(\alpha)}(x)|^2
\:.\ee

Obviously we have recovered in the particular case of graphs the
Onsager-Casimir relation relating injectivities and emissivities
\cite{But93}~:
\be
\rho(\alpha,x;\{\theta_{\mu\nu}\}) = \rho(x,\alpha;\{-\theta_{\mu\nu}\})
\:.\ee
Injectivities and emissivities are related by reversing the sign of the
magnetic field.

\noindent{\bf LDoS.}
If we sum the injectivities or the emissivities, we get the same result,
the contribution of the continuous part of the spectrum to the LDoS of the
graph~:
\be
\sum_\alpha \rho(\alpha,x) = \sum_\alpha \rho(x,\alpha)
= -\frac{1}{2\I\pi}\tr{ \Sigma^\dagger \frac{\delta\,\Sigma}{\delta V(x)}
}
\ee
From (\ref{inj2}) we demonstrate that~:
\be
-\frac{1}{2\I\pi}\frac{\delta}{\delta V(x)}\ln\det\Sigma
= \sum_\alpha |\tilde\psi^{(\alpha)}(x)|^2
\:.\ee

Once again, we insist that one of the interest of graphs is the fact that
the functional derivatives involved in the injectivities, emissivities or
the
LDoS can be computed with algebraic calculations~:
\bea
&& -\frac{1}{2\I\pi}\frac{\delta}{\delta V(x)}\ln\det\Sigma \nonumber\\
&=& \tr{\tilde\Psi_x^\dagger K_x\tilde\Psi_x}
= \frac1{\pi k}\tr{ W_x\frac{1}{-M_x+W_x^{\rm T}W_x} K_x
                    \frac{1}{M_x+W_x^{\rm T}W_x}W_x^{\rm T} } \\
&=& \tr{\tilde\Psi^\dagger K(x)K(x)^\dagger\tilde\Psi}
= \frac1{\pi k}\tr{ W\frac{1}{-M+W^{\rm T}W} K(x)K(x)^\dagger
                    \frac{1}{M+W^{\rm T}W}W^{\rm T} }
\:.\eea

\mathversion{bold}

\subsection*{Relation between $\D\Sigma/\D E$
             and $\delta\,\Sigma/\delta V(x)$}
\mathversion{normal}

If the relation (\ref{important}) is integrated over the whole graph, we
get the following exact relation
(see \cite{GasChrBut96} for the one-dimensional case and
the demonstration in an appendix of \cite{TexDeg02} for graphs)~:
\be
-\int_{\rm Graph}\D x\,
  \Sigma^\dagger\frac{\delta\,\Sigma}{\delta V(x)}
= \Sigma^\dagger\frac{\D\,\Sigma}{\D E}
+ \frac{1}{4E}\left(\Sigma-\Sigma^\dagger\right)
\:.\ee
If we trace this relation between matrices we obtain~:
\be
-\int_{\rm Graph}\D x\,\frac{\delta}{\delta V(x)}\ln\det\Sigma
=\frac{\D}{\D E} \ln\det\Sigma + \frac{1}{4E}\tr{\Sigma-\Sigma^\dagger}
\label{rel2}
\:.\ee
Note that in the usual formulation of the Friedel sum rule 
\cite{Fri52,Kre53,Hua63,LanLif66e} only the first term
appears. This is due to the fact that what is considered in this case is the 
variation of the total density of states (the LDoS integrated over the whole
space) due to the introduction of a potential,
whereas what we consider in the above formulae is the LDoS integrated in the
scattering region only (the graph). This point is discussed in detail in the 
appendix B of \cite{Tex02}.


\subsection*{Example~: The ring}

We apply in this section the above considerations to
a ring \cite{GefImrAzb84,ButImrAzb84}
of perimeter $l$ threaded by an Aharonov-Bohm flux $\theta$. 
The upper arm (arc $a$) is of length
$l_a$ and the lower arm (arc $b$) of length $l_b$ (we have $l=l_a+l_b$).
We consider the situation without potential for simplicity.
The ring is coupled to two leads at the two vertices $1$ and $2$ by
arbitrary
couplings described by the two real parameters $w_1$ and
$w_2$ (The maximum coupling is $w_i=1$~; in this case the scattering at
the
vertex is symmetric. The decoupling of the ring occurs for $w_i\to0$). We
call $\theta_{a,b}$ the line integral of the vector potential due to the
flux
along the two arcs~:
$\theta=\theta_a+\theta_b$.

\begin{figure}[!ht]
\begin{center}
\includegraphics[scale=1]{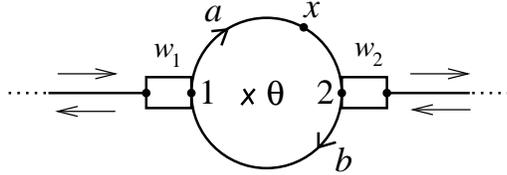}
\end{center}
\caption{Ring penetrated by an Aharonov-Bohm flux coupled to two leads. 
         The functional derivative of the scattering matrix at $x$ determines
         the local quantities: partial density of states, injectivities,
         emissivities and the local density of states.\label{ring1}}
\end{figure}

\noindent
The matrix $M$ of the ring is~:
\be
M =
\left(\begin{array}{cc}
\I cot_a+\I cot_b
  & -\frac{\I\EXP{-\I\theta_a}}{s_a}-\frac{\I\EXP{\I\theta_b}}{s_b}
\\[0.2cm]
-\frac{\I\EXP{\I\theta_a}}{s_a}-\frac{\I\EXP{-\I\theta_b}}{s_b}
  & \I cot_a+\I cot_b
\end{array}\right)
\ee
where $cot_{a,b}\equiv\cot kl_{a,b}$ and $s_{a,b}\equiv\sin kl_{a,b}$.
The matrix $W$ is~:
\be
W =
\left(\begin{array}{cc}
w_1 & 0 \\
0   & w_2
\end{array}\right)
\:.\ee
From (\ref{RES2}) we obtain~:
\be
\Sigma = -1 + \frac{2}{\tilde S}
\left(\begin{array}{cc}
  \I w_1^2\sin kl + w_1^2w_2^2 s_a s_b
& \I w_1 w_2 ( s_b\EXP{-\I\theta_a} + s_a\EXP{\I\theta_b} ) \\[0.2cm]
  \I w_2 w_1 ( s_b\EXP{\I\theta_a}  + s_a\EXP{-\I\theta_b} )
& \I w_2^2\sin kl + w_1^2w_2^2 s_a s_b
\end{array}\right)
\ee
where
\be
\tilde S = s_as_b\det(M+W^{\rm T}W)
= 2(\cos\theta-\cos kl) + \I(w_1^2+w_2^2)\sin kl + w_1^2w_2^2 s_a s_b
\ee
is the modified spectral determinant (the spectral determinant $S(-E)$
is obtained by taking $w_{1,2}=0$).

\begin{figure}[!ht]
\begin{center}
\includegraphics{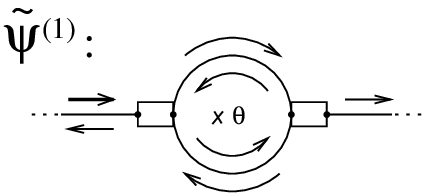}
\hspace{2cm}
\includegraphics{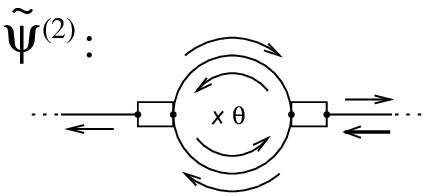}
\end{center}
\caption{Scattering states incident from the left and the right.}
\end{figure}

The two components of the stationary state $(1)$ wave function on the
two arcs read~:
\bea
\tilde\psi_a^{(1)}(x) &=& \frac{\I w_1}{\sqrt{\pi k}\,\tilde S}
\EXP{\I \theta_a x/l_a}
\left[ \sin k(l-x) + \EXP{-\I\theta}\sin kx -\I w_2^2 s_b \sin k(l_a-x)
\right]\\
\tilde\psi_{\bar b}^{(1)}(x) &=& \frac{\I w_1}{\sqrt{\pi k}\,\tilde S}
\EXP{-\I \theta_b x/l_b}
\left[ \sin k(l-x) + \EXP{\I\theta}\sin kx -\I w_2^2 s_a \sin k(l_b-x)
\right]
\eea
and the stationary state $(2)$~:
\bea
\tilde\psi_a^{(2)}(x) &=& \frac{\I w_2}{\sqrt{\pi k}\,\tilde S}
\EXP{\I \theta_a (x/l_a-1)}
\left[ \sin k(l-l_a+x) + \EXP{\I\theta}\sin k(l_a-x) -\I w_1^2 s_b \sin kx
\right]\\
\tilde\psi_{\bar b}^{(2)}(x) &=& \frac{\I w_2}{\sqrt{\pi k}\,\tilde S}
\EXP{\I \theta_b (1-x/l_b)}
\left[ \sin k(l-l_b+x) + \EXP{-\I\theta}\sin k(l_b-x) -\I w_1^2 s_a \sin
kx
\right]
\:.\eea
From the choice of orientations of the arcs $a$ and $b$ (see figure) we
see
that for the two components, the coordinate $x$ measures the distance from
the
vertex $1$
(this is why we consider the component $\psi_{\bar b}(x)$ instead of
$\psi_{b}(x)$).
Then $\psi_a^{(\alpha)}(0)=\psi_{\bar b}^{(\alpha)}(0)$ and
$\psi_a^{(\alpha)}(l_a)=\psi_{\bar b}^{(\alpha)}(l_b)$.

Now we show how the local FSR is applied.
The matrix $M_x$ describes the same graph
with an additional vertex $x$ on arc $a$~:
\be
M_x =
\left(\begin{array}{ccc}
\I cot_{1x}+\I cot_b
  & -\frac{\I}{s_b} \EXP{\I\theta_b}
  & -\frac{\I}{s_{1x}} \EXP{-\I\theta_{x1}}\\[0.2cm]
-\frac{\I}{s_b} \EXP{-\I\theta_b}
  & \I cot_{2x}+\I cot_b
  & -\frac{\I}{s_{2x}} \EXP{\I\theta_{2x}}\\[0.2cm]
-\frac{\I}{s_{1x}} \EXP{\I\theta_{x1}}
  & -\frac{\I}{s_{2x}} \EXP{-\I\theta_{2x}}
  & \I cot_{1x}+\I cot_{2x}
\end{array}\right)
\ee
in the basis of vertices $\{1,2,x\}$. The notations $1x$ and $x2$
designate
the two arcs replacing the arc $a$. In particular~:
$l=l_{1x}+l_{2x}+l_b$ and $\theta=\theta_b+\theta_{2x}+\theta_{x1}$.
The coupling matrix is~:
\be
W_x=
\left(\begin{array}{ccc}
w_1 & 0   & 0 \\
0   & w_2 & 0
\end{array}\right)
\ee
and
\be
K_x=
\left(\begin{array}{ccc}
0 & 0 & 0 \\
0 & 0 & 0 \\
0 & 0 & 1
\end{array}\right)
\:.\ee
After a little of algebra, we can check that (\ref{fd1}) gives
(\ref{fd4}).
The wave function of the time-reversed graph is obtained by changing the
signs of all fluxes~:
$\psi^{t.r.(\alpha)}(x) = \psi^{(\alpha)}(x)\big|_{\theta_i\to-\theta_i}$.

A more direct derivation follows from (\ref{fd5}).
We can use more efficiently (\ref{fd5}). $x$ belongs to the arc $a$, then
we have~:
\be
K(x)K(x)^\dagger = \frac1{s_a^2}
\left(\begin{array}{c}
  \sin k(l_a-x) \\ \EXP{\I\theta_a}\sin kx
\end{array}\right)
\left(\begin{array}{cc}
  \sin k(l_a-x) & \EXP{-\I\theta_a}\sin kx
\end{array}\right)
\:.\ee
Using (\ref{fd5}) we can check easily that
\be
-\frac1{2\I\pi}\frac{\delta\,\Sigma}{\delta V(x)} =
\left(\begin{array}{c}
  \tilde\psi^{t.r.\,(1)}_{a}(x) \\  \tilde\psi^{t.r.\,(2)}_{a}(x)
\end{array}\right)
\left(\begin{array}{cc}
  \tilde\psi^{(1)}_{a}(x) & \tilde\psi^{(2)}_{a}(x)
\end{array}\right)
\ee
Together with $\Sigma$ this permits to calculate the density of states
of interest.

\subsection*{A ring out-of-equilibrium studied by STM}

We now give a physical application~: a ring
probed by scanning tunneling microscopy (STM).

\begin{figure}[!ht]
\begin{center}
\includegraphics[scale=1.25]{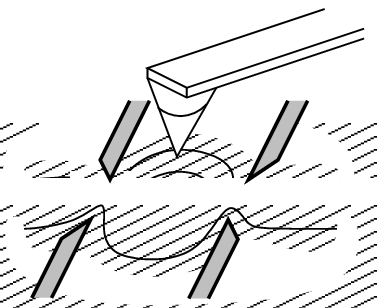}
\hspace{1cm}
\includegraphics{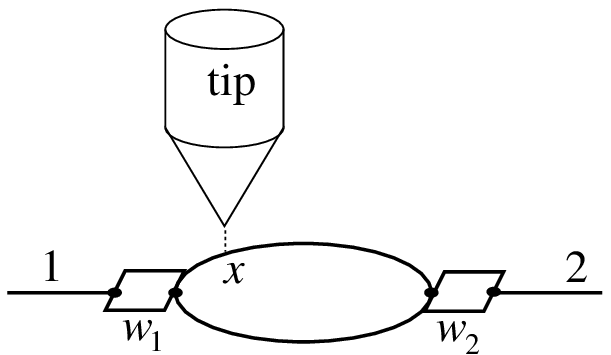}
\end{center}
\caption{{\it Left figure}~: Ahronov-Bohm ring probed by an STM. 
         The dashed region is forbidden to the electron gas at the
         interface of 2 semiconductors. The couplings to the contacts 
         can be adjusted with the help of gate voltages. 
        {\it Right figure}~: the graph that models this situation.}
\end{figure}

It is well known that if the system is at equilibrium, the tunneling
current
is related to the LDoS through the Bardeen formula.
If now the device under consideration is put in an out-of-equilibrium
situation by taking different potentials at the various contacts, which
induces current flow, the Bardeen formula has to be generalized.
We have to consider the conductances between the contact $\alpha$
and the STM tip. According to Refs. \cite{GraBut97,But02} the
corresponding transmission probabilities are
\be
T_{{\rm tip},\alpha} = 4\pi^2\rho_{\rm tip} |t_{\rm tip}|^2\rho(x,\alpha)
\:.\ee
Here $\rho_{\rm tip}$ is the DoS in the tip,
$t_{\rm tip}$ the transmission amplitude between the system and the tip
and $\rho(x,\alpha)$ is the injectivity from contact $\alpha$ into the
point $x$.

In the case of the ring threaded by a flux $\theta$, the injectivities 
were calculated above.
Transmission from the tip into contact $\alpha$ are related to the
emissivities~:
\be
T_{\alpha,{\rm tip}} = 4\pi^2\rho_{\rm tip} |t_{\rm tip}|^2\rho(\alpha,x)
\ee
whereas the modification of the conductances
(transmission probabilities) of the system due to the
presence of the tip involves the partial LDoS~:
\be
T^{{\rm tip}}_\ab = T_\ab -
4\pi^2\rho_{\rm tip} |t_{\rm tip}|^2\rho(\alpha,x,\beta)
\ee
where $T_\ab$ is the transmission amplitude from lead $\beta$ to $\alpha$
in the absence of the tip.

We consider in detail the case where the ring is also weakly
coupled to the two leads~: $w_{1,2}\ll1$. Transmission
sepctroscopy of a weakly coupled ring was already investigated in
Ref. \cite{ButImrAzb84}.
In this case the scattering matrix presents sharp resonant
(Breit-Wigner or Fano) structures. It is shown in \cite{TexDeg02} that,
if the spectrum is non
degenerate, the modulus of the stationary scattering state near an energy
$E_n$ of the isolated graph is
\be
\rho(x,\alpha)=
|\tilde\psi^{(\alpha)}_E(x)|^2\APPROX{E\sim E_n}
\frac1\pi \frac{\Gamma_{n,\alpha}}{(E-E_n)^2+\Gamma_n^2}\,|\varphi_n(x)|^2
\ee
where $\varphi_n(x)$ is the eigenstate of the isolated graph.
$\Gamma_{n,\alpha}=\sqrt{E_n}w_\alpha^2|\varphi_n(\alpha)|^2$ is the
contribution of the coupling to the lead $\alpha$ to the resonance width
$\Gamma_n=\sum_\alpha\Gamma_{n,\alpha}$. From the appendix \ref{app:TRS}
we also see that the emissivity is equal to the injectivity in the case
of the ring
$\rho(x,\alpha)=\rho(\alpha,x)$, then the conductances are equal~:
$T_{{\rm tip},\alpha}=T_{\alpha,{\rm tip}}$. [Note that in general
these two transmissions are not equal but 
$T_{{\rm tip},\alpha}(\theta)=T_{\alpha,{\rm tip}}(-\theta)$].
Using the equations of
appendix \ref{app:TRS} and the definition of the PLDoS we show
\be
\rho(\alpha,x,\beta)\APPROX{E\sim E_n}
\frac1\pi
\frac{\delta_\ab\Gamma_{n,\alpha}[(E-E_n)^2-\Gamma_n^2]
     +2\Gamma_n\Gamma_{n,\alpha}\Gamma_{n,\beta}}
     {[(E-E_n)^2+\Gamma_n^2]^2}\,|\varphi_n(x)|^2
\:.\ee

For the perfect ring (no potential),
since the eigenstates of the isolated graph are
$\varphi_n(x)=\frac1{\sqrt{l}}\EXP{2n\I\pi x/l}$ with $n\in\ZZ$, none of
the conductances depend on the position $x$ of the tip and they only
present
a resonant structure as a function of the Fermi energy $E_F$~:
\be
T_{{\rm tip},\alpha} \propto
\sum_n\frac{\Gamma_{n,\alpha}}{(E_F-E_n)^2+\Gamma_n^2}
\ee
with $E_n=\left(\frac{2n\pi-\theta}{l}\right)^2$
and
$\Gamma_{n,\alpha}=\frac1l\sqrt{E_n}w_\alpha^2$. As a function of the
magnetic flux $\theta$, the level $E_n(\theta)$ shifts and the conductance
presents also a resonant structure as a function of the flux.
If we increase the couplings $w_{1,2}$ to the leads, the resonance peaks
are broadened and oscillations in the injectivities are generated,
and thus an $x$-dependence of $T_{{\rm tip},\alpha}$, since the original
eigenstates of the isolated ring become strongly perturbed by the
couplings
to the leads.


\mathversion{bold}
\section{Functional derivative of $\Sigma$ in the arc language}
\mathversion{normal}

We have demonstrated above how the functional derivative of the scattering
matrix with respect to the potential is calculated. We have adopted a
vertex
point of view: all the matrices considered were matrices coupling the
vertices of the graph. The vertex language is rather efficient in the
sense
that it leads to the consideration of compact matrices of the smallest
possible size.
However this approach supposes that one can introduce vertex variables,
which can
be achieved only if the wave function is continuous at the vertices inside
the graph (the introduction of tunable couplings to the leads implies that
the wave function at the end of the lead is different from the wave
function
at the vertex, however the wave function is still continuous inside the
graph). The continuity of the wave function at the vertices implies a
particular choice for the scattering at the vertices: the transmission
amplitudes between all the leads issuing from the same vertex are
equal, a description that may not be absolutely satisfactory in all cases.
To describe the most general situation one has to abandon the constraint
of
continuity of the wave function at the vertices. In this case it is not
anymore possible to define vertex variables and one has to introduce arc
variables. We recall here some notions presented in \cite{TexMon01}.
On each arc $i$ we introduce an amplitude $A_i$ arriving at
the vertex from which $i$ issues and an amplitude $B_i$ departing from it
(see figure). That is to say that the wave function $\psi_i(x)$ on the
bond is matched with $A_i\EXP{-\I kx}+B_i\EXP{\I kx}$ at the extremity
of the arc.

\begin{figure}[!ht]
\begin{center}
\includegraphics{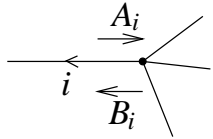}
\end{center}
\caption{The two amplitudes associated to the arc $i$.\label{arci}}
\end{figure}

\noindent
It is clear that we have to introduce $L$ such couples of amplitudes, one
for each external lead. These external amplitudes are gathered in
$L$-column
vectors $A^{\rm ext}$ and $B^{\rm ext}$. By definition the scattering
matrix
relates these amplitudes~: $B^{\rm ext}=\Sigma A^{\rm ext}$.
On the other hand we must introduce two couples of amplitudes $A_i$, $B_i$
per bond of the graph, {\it i.e.} one couple per arc. We gather these $2B$
amplitudes into the column vectors $A^{\rm int}$ and $B^{\rm int}$.
Finally we group all amplitudes, internal and external, in two
$2B+L$ column vectors $A$ and $B$.

The scattering by the bonds is described by a matrix
$R$ coupling reversed internal arcs~: $A^{\rm int}=RB^{\rm int}$.
The scattering at the vertices is described by a matrix $Q$ coupling
arcs issuing from the same vertex~: $B=QA$.
If the basis of arcs is organized as
$\{\mbox{internal arcs},\mbox{external arcs}\}$, the matrix $Q$ is
separated
into blocks:
\be\label{vs}
Q =
\left(
\begin{array}{c|c}
Q^{\rm int} & \tilde Q^{\rm T} \\ \hline
\tilde Q   & Q^{\rm ext}
\end{array}
\right)
\ee
For simplicity we
suppose here that the vertex scattering matrix is symmetric, that is the
scattering at the vertices is not influenced by the presence of a magnetic
field. However it is straightforward to extend the results to the more
general situation: one has to make a disctinction between the two
off-diagonal
blocks of $Q$.
The scattering matrix reads \cite{TexMon01}:
\be\label{RES1}
\Sigma = Q^{\rm ext} +
\tilde Q \, ({R^{\dagger} - Q^{\rm int}})^{-1} \, \tilde Q^{\rm T}
\:.\ee

We now follow the same methodology as in the vertex approach to derive an
expression of the functional derivative of $\Sigma$ involving arc
matrices.
We consider a new graph ${\cal G}^\lambda$ which is similar to the
original
graph ${\cal G}$, apart from the fact
that it possesses an additional vertex at $x$ of weight
$\lambda$ (we recall that ${\cal G}$ and ${\cal G}_x={\cal G}^{\lambda=0}$
possess the same properties).
We call $R_x$ the new bond scattering matrix. It describes the same
physics
as $R$. The vertex scattering matrix is $Q^\lambda$. Compared to $Q$,
$Q^\lambda$ has an additional $2\times2$ block describing the scattering
at
the vertex $x$. This block couples the two arcs issuing from $x$~:
\be
\frac{2}{2+\I k/\lambda}
\left(\begin{array}{cc}1&1\\1&1\end{array}\right) - 1
\:.\ee
We introduce the notation $Q_x=Q^{\lambda=0}$. This matrix describes the
same physical situation as $Q$.
If we expand $Q^\lambda$ in powers of $\lambda$ we get for the internal
part, at first order~:
\be
(Q^\lambda)^{\rm int}=Q_x^{\rm int} - \frac{\I \lambda}{2k}
\kappa_x+\cdots
\:\ee
($\kappa_x$ couples the internal arcs). The only non vanishing elements
are those associated to the two arcs issuing from $x$:
\be
\kappa_x =
\left(\begin{array}{cc|cc|cc}
\ddots & \vdots & \vdots & \vdots & \vdots & \\
\cdots & 0      & 0      & 0      & 0      & \cdots \\ \hline
\cdots & 0      & 1      & 1      & 0      & \cdots \\
\cdots & 0      & 1      & 1      & 0      & \cdots \\ \hline
\cdots & 0      & 0      & 0      & 0      & \cdots \\
       & \vdots & \vdots & \vdots & \vdots & \ddots
\end{array}\right)
=
\left(\begin{array}{c}
\vdots\\0\\\hline 1\\1\\\hline0\\\vdots
\end{array}\right)
\left(\begin{array}{cc|cc|cc}
\cdots & 0      & 1      & 1      & 0      & \cdots
\end{array}\right)
\:.\ee
The functional derivative of $\Sigma$ is equal to the first order term of
$\Sigma^\lambda$. We get~:
\be\label{fda}
\frac{\delta\,\Sigma}{\delta V(x)}
=-\frac{\I}{2k}
\tilde Q_x (R_x^\dagger - Q_x^{\rm int})^{-1} \kappa_x
(R_x^\dagger - Q_x^{\rm int})^{-1}\tilde Q_x^{\rm T}
\:.\ee
This expression allows to compute the functional derivative of $\Sigma$
in terms of matrices describing the graph ${\cal G}_x$.
The component of the stationary scattering state on the
bond $(\ab)$ to which $x$ belongs is given by
\be
\sqrt{4\pi k}\:\tilde\psi^{(\mu)}_{(\ab)}(x)
=\left[\left(\begin{array}{cccccc}
\cdots & 0      & 1      & 1      & 0      & \cdots
\end{array}\right)
(R_x^\dagger - Q_x^{\rm int})^{-1}\tilde Q_x^{\rm T}
\right]_\mu
\:.\ee
With these results we can recover the relation (\ref{fd4}) within the arc
approach.

We are now going to simplify this expression in the sense that we will
relate the functional derivative to the matrices of the original graph
${\cal G}$, of smaller size than those of ${\cal G}_x$.

We suppose that $x$ belongs to the bond $(\alpha\beta)$ of ${\cal G}$.
We show the structure of the matrices describing ${\cal G}$ in a basis
$\{\mbox{other internal arcs},\ab,\ba\,||\mbox{external arcs}\}$.
The vertex scattering matrix is given by
\be
Q =
\left(
\begin{array}{c||c}
Q^{\rm int} & \tilde Q^{\rm T} \\ \hline\hline
\tilde Q   & Q^{\rm ext}
\end{array}
\right)
\ee
where we separate by a double line the internal and external vertices.
The bond scattering matrix is
\be
R =
\left(\begin{array}{cc}
R_{\rm oia} & 0 \\
0            &
\left(\begin{array}{cc}r_\ab&t_\ba\\t_\ab&r_\ba\end{array}\right)
\end{array}\right)
\ee
where $R_{\rm oia}$ is the block coupling all internal arcs apart from
$\ab$ and $\ba$.

We now examine the structure of the matrices describing the graph
${\cal G}_x$.
The bond $(\alpha\beta)$ of ${\cal G}$ is replaced by two bonds
$(\alpha x)$ and $(x\beta)$ in ${\cal G}_x$.

\begin{figure}[!ht]
\begin{center}
\includegraphics{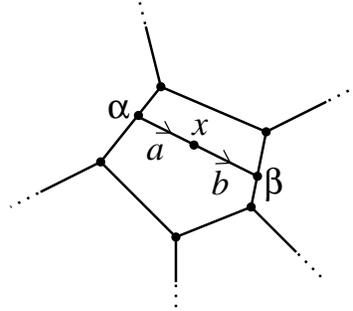}
\end{center}
\caption{Auxiliary bonds $a\equiv\alpha x$ and $b\equiv\beta x$ on the 
         $\alpha \beta$.\label{aux}}
\end{figure}

The two arcs will be also denoted by~: $a\equiv\alpha x$ and $b\equiv x\beta$
(see Fig. \ref{aux}).
For simplicity we chose to organize the basis of arcs as~:
$\{\mbox{other internal arcs},a,\bar b\,|\,\bar a,b\,||\mbox{external arcs}\}$.
The vertex scattering matrix reads~:
\be
Q_x=
\left(\begin{array}{c|c||c}
Q^{\rm int} & 0     & \tilde Q^{\rm T} \\ \hline
0           & \left(\begin{array}{cc}0&1\\1&0\end{array}\right)
                    & 0 \\ \hline\hline
\tilde Q    & 0     & Q^{\rm ext}
\end{array}\right)
\ee
where the blocks of $Q$ appear. We separate by a single line the part
reminiscent of ${\cal G}$ and the one added by considering
${\cal G}_x$.
The bond scattering matrix is
\be
R_x =
\left(\begin{array}{cc|c}
R_{\rm oia} & 0 & 0\\
0
  & \left(\begin{array}{cc}r_a&0\\0&r_{\bar b}\end{array}\right)
  & \left(\begin{array}{cc}t_{\bar a}&0\\0&t_b\end{array}\right)\\
    [0.4cm]\hline
0
  & \left(\begin{array}{cc}t_a&0\\0&t_{\bar b}\end{array}\right)
  & \left(\begin{array}{cc}r_{\bar a}&0\\0&r_b\end{array}\right)
\end{array}\right)
\:.\ee
Note that the transmission and reflexion coefficients of the arcs
$a=\alpha x$
and $b=x\beta$ are related to the ones of the arc $\alpha\beta$ through
the
following relation~:
\be
\left(\begin{array}{cc}r_\ab&t_\ba\\t_\ab&r_\ba\end{array}\right)^{-1}
=
 \left(\begin{array}{cc}r_a^*&0\\0&r_{\bar b}^*\end{array}\right)
-\left(\begin{array}{cc}t_a^*&0\\0&t_{\bar b}^*\end{array}\right)
 \left(\begin{array}{cc}r_{\bar a}^*&-1\\-1&r_b^*\end{array}\right)^{-1}
 \left(\begin{array}{cc}t_{\bar a}^*&0\\0&t_b^*\end{array}\right)
\:.\ee
It implies the expected relations~:
$r_\ab=r_a+\frac{t_{\bar a}r_bt_a}{1-r_{\bar a}r_b}$,
$t_\ab=\frac{t_at_b}{1-r_{\bar a}r_b}$, etc.
(A remark about notations~: $t_\ab$ is the transmission amplitude from
$\alpha$ to $\beta$ since it is the transmission of the arc $\ab$
connecting $\alpha$ to $\beta$.)
To compute the functional derivative of the scattering matrix, we need to
find the inverse of
\be
R_x^\dagger-Q_x^{\rm int}
=\left(\begin{array}{c|c}
\cdots
  & \begin{array}{c}0\\
       \left(\begin{array}{cc}t_a^*&0\\0&t_{\bar b}^*\end{array}\right)
    \end{array} \\ \hline
\begin{array}{cc}0&
  \left(\begin{array}{cc}t_{\bar a}^*&0\\0&t_b^*\end{array}\right)
\end{array}
  & \left(\begin{array}{cc}r_{\bar a}^*&-1\\-1&r_b^*\end{array}\right)
\end{array}\right)
=\left(\begin{array}{c|c} A&B\\ \hline C&D\end{array}\right)
\:.\ee
One can see that
\be
A - B D^{-1} C = R^\dagger-Q^{\rm int}
\ee
involves the matrices of ${\cal G}$. We have~:
\be
(R_x^\dagger-Q_x^{\rm int})^{-1}
=\left(\begin{array}{c|c}
(R^\dagger-Q^{\rm int})^{-1} & -(R^\dagger-Q^{\rm int})^{-1}B D^{-1} \\
\hline
-D^{-1} C(R^\dagger-Q^{\rm int})^{-1} & \cdots
\end{array}\right)
\:.\ee
Then after a little bit of algebra, (\ref{fda}) gives~:
\be\label{fda2}
\frac{\delta\,\Sigma}{\delta V(x)}
=-\frac{\I}{2k}
\tilde Q (R^\dagger - Q^{\rm int})^{-1} {\cal L}(x)
(R^\dagger - Q^{\rm int})^{-1}\tilde Q^{\rm T}
\ee
where we have introduced the matrix ${\cal L}(x)$, that couples only
the two arcs $\alpha\beta$ on which is put the vertex
$x$~:
\bea
{\cal L}(x)
=\left(\begin{array}{c}
\vdots\\0\\
      \frac{t_{\alpha x}^*(1+r_{x\beta}^*)}{1-r_{x\alpha}^*r_{x\beta}^*}\\
      \frac{t_{\beta x}^*(1+r_{x\alpha}^*)}{1-r_{x\alpha}^*r_{x\beta}^*}
\end{array}\right)
\left(\begin{array}{cccc}
\cdots&0&
      \frac{(1+r_{x\beta}^*) t_{x\alpha}^*}{1-r_{x\alpha}^*r_{x\beta}^*}&
      \frac{(1+r_{x\alpha}^*) t_{x\beta}^*}{1-r_{x\alpha}^*r_{x\beta}^*}
\end{array}\right)
\:.\eea
We have now achieved our program since (\ref{fda2}) involves matrices of the
original graph ${\cal G}$, and not bigger matrices of the graph ${\cal G}_x$
(describing however the same situation as ${\cal G}$) with
an additional vertex at $x$.

Note that in the free case (no potential) the transmissions are
$t_{\mu\nu}=\EXP{\I kl_{\mu\nu}+\I\theta_{\mu\nu}}$ and the above matrix
takes the simple form~:
\be
{\cal L}(x)=
\left(\begin{array}{cccc}
\ddots & \vdots  & \vdots        & \vdots \\
\cdots & 0 & 0             & 0      \\
\cdots & 0 & \EXP{-2\I kx_\ab} & \EXP{-\I kl_\ab-\I\theta_\ab}\\
\cdots & 0 & \EXP{-\I kl_\ab+\I\theta_\ab} & \EXP{-2\I kx_\ba}
\end{array}\right)
\:.\ee

Following what has been done previously we try to obtain a more
transparent
expression of ${\cal L}(x)$.
The starting point is to express the wave function on a given bond $(\ab)$
in terms of the arc amplitudes. In the arc language, the appropriate basis
of solutions of the Schr\"odinger equation on the bond
$[E+\D_x^2 - V_{(\ab)}(x)]f(x)=0$ is not anymore the functions
$f_\ab(x)$ and $f_\ba(x)$, but the couple of stationary scattering states
$\phi_\ab(x)$ and $\phi_\ba(x)$ associated with the potential
$V_{(\ab)}(x)$
on the bond $(\ab)$.
The function $\phi_\ab(x)$ is the scattering state incoming on the bond
from the vertex $\alpha$ and is matched out of the bond like~:
$\phi_\ab(x)=\EXP{\I kx}+r_\ab\EXP{-\I kx}$ for $x<0$ and
$\phi_\ab(x)=t_\ab\EXP{\I k(x-l_\ab)}$ for $x>l_\ab$ \cite{TexMon01}.

\begin{figure}[!ht]
\begin{center}
\includegraphics{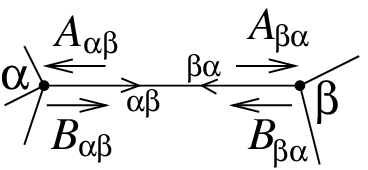}
\end{center}
\caption{The two couples of amplitudes $A_{\ab},\,B_{\ab}$ 
         and $A_{\ba},\,B_{\ba}$ related to the arcs $\ab$ and $\ba$.}
\end{figure}

Then the component of the wave function on the bond reads~:
\be
\psi_{(\ab)}(x_\ab) = B_\ab\,\phi_\ab(x_\ab)+B_\ba\,\phi_\ba(x_\ab)
\:.\ee
If we introduce the $2B$-vector
\be
{\cal K}(x)^\dagger =
\left(\begin{array}{cccc}
\cdots & 0 & \phi_\ab(x_\ab) & \phi_\ba(x_\ab)
\end{array}\right)
\ee
we can write
\be
\psi_{(\ab)}(x_\ab) = {\cal K}(x)^\dagger B^{\rm int}
\:.\ee

We have now to specify the value of the amplitudes $B_\ab$ and $B_\ba$ for
the stationary scattering state. If we consider the scattering state
$\tilde\psi^{(\mu)}$, it is described by external incoming amplitudes
$\tilde A^{(\mu){\rm ext}}$ whose components
$\tilde A^{(\mu){\rm ext}}_\nu=\frac{1}{\sqrt{4\pi k}}\delta_{\mu\nu}$
describe a plane wave entering the graph from the lead connected to the vertex
$\mu$.
From the two equations $A^{\rm int}=RB^{\rm int}$ and $B=QA$ we get
\be
\tilde B^{(\mu)\rm int} =
(1-Q^{\rm int}R)^{-1}\tilde Q^{\rm T}\tilde A^{(\mu){\rm ext}}
\ee
then
\be
\tilde \psi^{(\mu)}_{(\ab)}(x_\ab) = {\cal K}(x)^\dagger
(1-Q^{\rm int}R)^{-1}\tilde Q^{\rm T}\tilde A^{(\mu){\rm ext}}
=\frac{1}{\sqrt{4\pi k}}
\left[
{\cal K}(x)^\dagger (1-Q^{\rm int}R)^{-1}\tilde Q^{\rm T}
\right]_\mu
\:.\ee
Therefore, using (\ref{fd4}) we obtain~:
\be\label{fda3}
-\frac{1}{2\I\pi}\frac{\delta\,\Sigma}{\delta V(x)}
=\frac{1}{4\pi k}
\tilde Q (1- RQ^{\rm int})^{-1} R{\cal K}(x){\cal K}(x)^\dagger
(1- Q^{\rm int}R)^{-1}\tilde Q^{\rm T}
\ee
which shows that
\be
{\cal L}(x)R={\cal K}(x){\cal K}(x)^\dagger
\:.\ee
We have derived here non trivial relations between the reflection and
transmission coefficients for the bonds $(\alpha x)$ and $(\beta x)$
on one hand, and the functions $\phi_\ab(x)$ and $\phi_\ba(x)$ on the
other hand.

\subsection{Injectivities and emissivities}

To compute the objects involved in the injectivities and emissivities, one
needs to use the following relations~:
\be
\Sigma^\dagger\,\tilde Q \, ({R^{\dagger} - Q^{\rm int}})^{-1}
= \tilde Q^* \, (R - Q^{\rm int\:\dagger})^{-1} R
\ee
and
\be
({R^{\dagger} - Q^{\rm int}})^{-1} \, \tilde Q^{\rm T}\,\Sigma^\dagger
=R\,(R - Q^{\rm int\:\dagger})^{-1}\,\tilde Q^\dagger
\:.\ee
The demonstration of these two relations uses (\ref{RES1}) and relations
between the blocks of $Q$ coming from its unitarity.
The computation of injectivities requires an expression for~:
\be
-\frac{1}{2\I\pi}\Sigma^\dagger \frac{\delta\,\Sigma}{\delta V(x)}
=\frac{1}{4\pi k}
\tilde Q^* \, (1 - R^\dagger Q^{\rm int\:\dagger})^{-1}
{\cal K}(x){\cal K}(x)^\dagger
(1 - Q^{\rm int}R)^{-1}\tilde Q^{\rm T}
\ee
and the emissivities require~:
\be
-\frac{1}{2\I\pi}\frac{\delta\,\Sigma}{\delta V(x)}\Sigma^\dagger
=\frac{1}{4\pi k}
\tilde Q (R^\dagger - Q^{\rm int})^{-1} {\cal K}(x){\cal K}(x)^\dagger
\,(R - Q^{\rm int\:\dagger})^{-1}\,\tilde Q^\dagger
\:.\ee

\subsection{Example~: the ring with one lead}

We consider the ring with only one lead.
This geometry has been used to illustrate the decoherence introduced
into a closed graph (here a ring) by the coupling to an external
lead \cite{But85}. We now apply to this geometry the arc approach.

\begin{figure}[!ht]
\begin{center}
\includegraphics{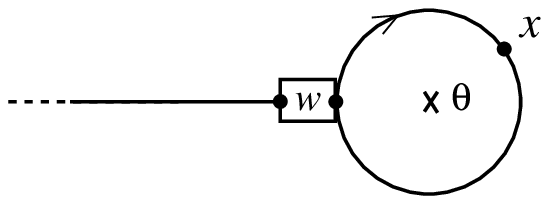}
\end{center}
\caption{}
\end{figure}

\noindent
We have~:
\be\label{Qring}
Q =
\frac{2}{2+w^2}
\left(\begin{array}{cc||c}
1 & 1 & w \\
1 & 1 & w \\ \hline\hline
w & w & w^2
\end{array}\right) -1
\ee
and
\be\label{Rring}
R =\left(\begin{array}{cc}
0 & \EXP{\I kl-\I\theta} \\
\EXP{\I kl+\I\theta} & 0
\end{array}\right)
\:.\ee
The matrix ${\cal L}(x)$ is~:
\be
{\cal L}(x)=
\left(\begin{array}{cc}
\EXP{-2\I kx} & \EXP{-\I kl-\I\theta}\\
\EXP{-\I kl+\I\theta} & \EXP{-2\I k(l-x)}
\end{array}\right)
\:.\ee

We have
$\det(R^\dagger-Q^{\rm int})=\frac2{2+w^2}\EXP{-\I kl}\,\tilde S$ where
the modified spectral determinant is
\be
\tilde S = 2(\cos\theta-\cos kl) + \I w^2\sin kl
\:.\ee
After a little bit of algebra we get, using (\ref{fda2})~:
\be
\frac{\delta\,\Sigma}{\delta V(x)}=
\frac{2\I w^2}{k\tilde S^2}
\left[\sin^2kl+2\sin kx\sin k(l-x)(\cos\theta-\cos kl)\right]
\:.\ee
We can check that this expression coincides with the one proven above
\be
\frac{\delta\,\Sigma}{\delta V(x)}
=-2\I\pi\,\tilde\psi^{t.r.}(x)\,\tilde\psi(x)
\ee
by using the wave function $\tilde\psi(x)$ that has been computed before
(it is given by the wave function of the ring with two leads
$\tilde\psi^{(1)}_a(x)$ in which we set $w_2=0$)~:
$\tilde\psi(x)=\frac{\I w}{\sqrt{\pi k}\,\tilde S}\EXP{\I\theta x/l}
(\sin k(l-x)+\EXP{-\I\theta}\sin kx)$.


\section{Conclusion}

In this article, we have discussed the local Friedel sum rule
which allows to extract local information (the local density of states,
injectivities, emissivities, and partial density of states)
from the scattering matrix. For this purpose one has to
be able to
compute functional derivatives of the scattering matrix, which can be
achieved for graphs by algebraic calculations.
We  have presented a
discussion both within the vertex language and the arc language. The former
is appropriate when we consider wave functions which are continuous at the
vertices
of the graph, whereas the latter is needed for the most general graphs
with arbitrary vertex scattering.
We have emphasized that the scattering can only give information about
the states of the graph coupled to the external leads. If the graph
possesses some states that remain uncoupled to the leads, this part of
the spectrum (discrete part) is not probed by elastic
single particle scattering.
Local information on density of states is important
for the solution of physical problems which involve a changing
charge distribution, like non-linear transport and
frequency dependent transport. Here we 
have given several examples
to demonstrate the applications of our formalism and in particular
have considered the experimental situation in which 
the local properties of a system are
probed by scanning tunneling microscopy.

The local density of states and its generalizations investigated
here are not the only quantities related to functional derivatives
of the scattering matrix. Off-diagonal elements of the functional
derivatives of the scattering matrix
are related to (dynamic) charge fluctuations \cite{But96}
which play an
important role in the description of dephasing in mesoscopic
conductors \cite{But00,PilBut02,CleGirSto02}. 
Furthermore low frequency parametric pumping
of electrical conductors \cite{Bro98,AvrElgGraSad01}
and the fluctuations associated with it \cite{MosBut02}
can be related to functional derivatives. Thus the work presented here
can be expected to be useful in a wide range of physical problems.

\section*{Acknowledgement} 

This work was supported by the Swiss National Science Foundation.  


\begin{appendix}

\section{Spectrum of open graphs - Continuous and discrete parts
         \label{app:spectrum}}

We have already mentioned in the introduction that certain graphs may
possess some localized states that are not coupled to the leads.
Since such states are not manifesting themselves in the scattering properties,
the usual state counting method from the scattering (Friedel sum rule) fails
\cite{Tex02}. 
The purpose of the appendix is to discuss the structure of the LDoS when 
such a situation occurs.

For each arc $i$ of a graph, we introduce an amplitude $A_i$ arriving at
the vertex from which $i$ issues, and an outgoing amplitude $B_i$ (figure
\ref{arci}). Those
amplitudes are related by vertex scattering $B=Q\,A$. The amplitudes
associated
with internal arcs are also related through the bond scattering matrix
$A^{\rm int}=R\,B^{\rm int}$. If we eliminate $B^{\rm int}$ we get
\bea\label{S1}
\tilde Q^{\rm T}\, A^{\rm ext}
0           &=& (R^\dagger-Q^{\rm int}) \, A^{\rm int} \\
\label{S2}
B^{\rm ext}&=& \tilde Q\, A^{\rm int} + Q^{\rm ext}\,A^{\rm ext}
\,.\eea
Two cases occur~:

\noindent({\it i}) In general $\det(R^\dagger-Q^{\rm int})\neq0$. Then at
any
energy $k^2$, the solution of the above equations has components on all
bonds of the graph and on the leads. All the solutions of the
Schr\"odinger
equation are the stationary scattering states.
In this case, the FSR gives the correct information on the number of
states of the graph\footnote{
  Note that $\det(R^\dagger-Q^{\rm int})\neq0$ for the one-dimensional
  case.
  The 1d case may be viewed as a single bond (two arcs $a$ and ${\bar a}$)
  connected at its extremities to two leads (arcs $1$ and $2$).
  The corresponding $4\times4$ vertex scattering matrix $Q$ is decomposed
  into the four $2\times2$ blocks which can always be chosen as~:
  $Q^{\rm ext}=Q^{\rm int}=0$ and $\tilde Q=\tilde Q^{\rm T}=1$.
  Then $|\det(R^\dagger-Q^{\rm int})|=1$. The FSR always counts correctly 
  the states in 1d, as it should.}.

\noindent({\it ii}) For certain graphs $\det(R^\dagger-Q^{\rm int})=0$
possesses a discrete set of solutions $E=E_1,E_2,\cdots$. At those
energies
one can construct solutions of the Schr\"odinger equation such that
$A^{\rm ext}=B^{\rm ext}=0$ while the internal amplitudes satisfy
$(R^\dagger-Q^{\rm int})A^{\rm int}=0$ and $\tilde Q A^{\rm int}=0$.
These relations describe a wave function localized inside the graph.
In this case, the continuous spectrum related with the stationary
scattering
states coexists with a discrete spectrum of localized states. The LDoS
takes
the form~:
\be
\rho(x;E) =
\hspace{-0.5cm}\underbrace{
  \sum_\alpha |\tilde\psi_E^{(\alpha)}(x)|^2
}_{\mbox{continuous spectrum}}\hspace{-0.5cm}
+
\underbrace{
 \sum_n \sum_{j=1}^{g_n} \delta(E-E_n)\,|\varphi_{n,j}(x)|^2
}_{\mbox{discrete spectrum}}
\ee
for $x$ in the graph (if $x$ belongs to a lead, the second part of course
vanishes). $\varphi_{n,j}(x)$ is a wave function localized in
the graph and normalized to unity and $j$ a degeneracy label.
The number of uncoupled states at energy $E_n$ is
$g_n={\rm dim}\,{\rm Ker}(R^\dagger-Q^{\rm int})|_{E=E_n}$.

We emphasize that what is unusual here is that the discrete and continous
parts of the spectrum coexist at the same energies, due to the absence of
hybridization of the localized states with the states of the continuous
spectrum.

\vspace{0.25cm}

The second situation may occur if the spectrum of the isolated graph is
degenerate. In the space of the parameters of the graph (lengths,
fluxes,...)
this occur in a volume of measure zero. This ``violation'' of the FSR
occurs
for discrete values of the parameters and signals a discontinuous
behaviour
of the scattering matrix as a function of those parameters.

\subsection{Example}

Let us consider the case of the ring coupled to one lead. This example
has already been studied in \cite{Tex02} in the vertex language. Here we
adopt an arc language needed to construct all the solutions of the
Schr\"odinger equation in the case ({\it ii}) and compute the LDoS.
The matrices $R$ and $Q$ are given above (\ref{Qring},\ref{Rring}).
\be
R^\dagger-Q^{\rm int} =\left(\begin{array}{cc}
\frac{w^2}{2+w^2} & \EXP{-\I kl-\I\theta}-\frac{2}{2+w^2}  \\
\EXP{-\I kl+\I\theta}-\frac{2}{2+w^2} & \frac{w^2}{2+w^2}
\end{array}\right)
\:,\ee
whose determinant is
\be
\det(R^\dagger-Q^{\rm int})=\frac2{2+w^2}\EXP{-\I kl}
\left[2(\cos\theta-\cos kl) + \I w^2\sin kl\right]
\:.\ee
\noindent({\it i}) If \mathversion{bold}$\theta\neq0$\mathversion{normal},
the equation
$\det(R^\dagger-Q^{\rm int})=0$ doesn't have a solution. The only solution
of the Schr\"odinger equation is the stationary scattering state. Then
the LDoS is
\be
\rho(x;E)= \left|\tilde\psi(x)\right|^2 =
  \left|
    \frac{w}{\sqrt{\pi k}}
    \frac{\sin k(l-x)+\EXP{-\I\theta}\sin kx}
         {2(\cos\theta-\cos kl) + \I w^2\sin kl}
  \right|^2
\:.\ee
In the limit $w\to0$, the wave function $\tilde\psi(x)$ gives the wave
functions of the isolated ring.

\noindent({\it ii}) \mathversion{bold}$\theta=0$~:\mathversion{normal}
However if the flux is zero (or a multiple of the flux quantum),
\be
\det(R^\dagger-Q^{\rm int})=\frac4{2+w^2}\EXP{-\I kl}
\left[2\sin(kl/2) + \I w^2\cos(kl/2)\right]\sin(kl/2)
\ee
vanishes for all energies of the isolated ring~: $k_n=2n\pi/l$ for
$n\in\NN$.
At $k\neq k_n$ the solutions of (\ref{S1},\ref{S2}) give the stationary
scattering state.
At $k=k_n$, (\ref{S1},\ref{S2}) possesses a solution
$A^{\rm int}=(1,1)\times A^{\rm ext}/w$ coupled to the lead (the
scattering
state), and a solution $A^{\rm int}=(1,-1)$ with $A^{\rm ext}=B^{\rm
ext}=0$
localized in the ring.
The LDoS is then~:
\be
\rho(x;E)=
  \left|
    \frac{w}{\sqrt{\pi k}}
    \frac{\cos k(x-l/2)}{2\sin(kl/2) + \I w^2\cos(kl/2)}
  \right|^2
+
  \sum_{n=1}^\infty \delta(E-k_n^2)
  \left|
    \sqrt{\frac2l} \sin k_nx
  \right|^2
\:.\ee
The first term is $-\frac{1}{2\I\pi}\frac{\delta}{\delta\,V(x)}\ln\Sigma$.

\vspace{0.25cm}

\mathversion{bold}
\noindent{\bf Continuity of $\Sigma$}
\mathversion{normal}

To end the section, let us discuss the question of continuity of $\Sigma$.
The scattering matrix should be a continuous function of the energy
obviously,
however it has no reason to be continuous as a function of parameters such
as
fluxes, lengths,...
In the case of the ring we have~:
\be
\Sigma(k^2,\theta)
=\EXP{\I\delta_f}=
\frac{\phantom{-}2(\cos kl-\cos\theta)+\I w^2\sin kl}
     {-2(\cos kl-\cos\theta)+\I w^2\sin kl}
\:.\ee
At zero flux~:
\be
\Sigma(k^2,0)
=
\frac{         - 2\sin(kl/2)+\I w^2\cos(kl/2)}
     {\phantom{-}2\sin(kl/2)+\I w^2\cos(kl/2)}
\:.\ee
When we study the system through its scattering properties, there is no
reason to introduce some arbitrary jumps of $\delta_f(k^2)$, and the only
natural choice is to impose the continuity of $\delta_f(k^2)$ as a
function
of energy, by convention.
Then, in the absence of localized states when the FSR holds,
it is related to the integrated DoS (IDoS) ${\cal N}(E)$ of the graph by
${\cal N}(E)\simeq\frac{1}{2\pi}\delta_f(E)$ up to some oscillatory term
inessential at the level of the Weyl term of the IDoS.

In the case of the ring we see that $\Sigma$ is discontinuous as a
function
of $\theta$. For example at the energies $kl=2n\pi$~:
\be
\lim_{\theta\to0}\lim_{kl\to2n\pi}\Sigma(k^2,\theta)=-1
\ee
whereas
\be
\lim_{kl\to2n\pi}\lim_{\theta\to0}\Sigma(k^2,\theta)=+1
\:.\ee
The discontinuous behaviour is even more stricking on the Friedel phase
which is plotted for different values of $\theta$. If $\theta\neq0$ the
Weyl term of the Friedel phase is $\delta_f\simeq2kl$, but if $\theta=0$
it only grows like $\delta_f\simeq kl$ \cite{Tex02} as illustrated on the
figure.

\begin{figure}[!ht]
\begin{center}
\includegraphics[scale=0.5]{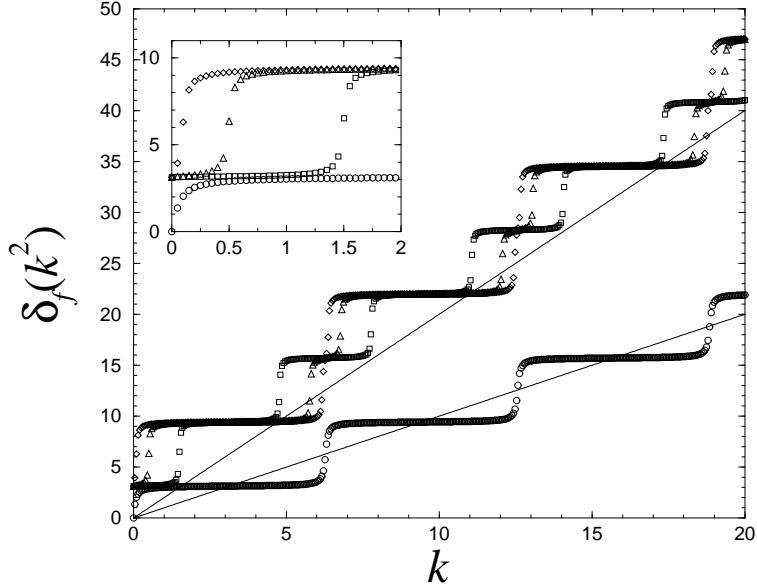}
\end{center}
\caption{Friedel phase $\delta_f(k^2)$. The coupling is $w=0.25$
         and the length $l=1$. The four
         curves corresponds to $\theta=1.5$ (squares), $\theta=0.5$
         (triangles), $\theta=0.1$ (diamonds) and $\theta=0$ (circles).
         The two lines are $kl$ and $2kl$. 
         If $\theta\neq0$ there are two jumps of $2\pi$ per interval
         $[2n\pi,2(n+1)\pi[$ at the energies of the two eigenstates of the
         isolated ring.
         If $\theta=0$ there is only one jump of $2\pi$ at the degenerate
         energy.
         In the inset we clearly see
         the discontinuity of $\delta_f(k^2=0)$ which equals $\pi$ for
         $\theta\neq0$ and $0$ for $\theta=0$.}
\end{figure}


\section{Time reversal symmetry and some useful relations \label{app:TRS}}

Note that the unitarity of (\ref{sm}) is demonstrated by using the
fact that the matrix $M$ is anti-hermitian \cite{TexMon01}
$M^\dagger = - M$ and by using the obvious relation~:
\be
 \frac{1}{M+W^{\rm T}W}
+\frac{1}{-M+W^{\rm T}W}
=\frac{1}{M+W^{\rm T}W}2W^{\rm T}W
 \frac{1}{-M+W^{\rm T}W}
\:.\ee

With the help of this relation we can easily find the relation between
the wave function and the one of the time-reversed graph. From
(\ref{sm},\ref{wf},\ref{trwf})~:
\be\label{ur1}
\Sigma^\dagger (\tilde\Psi^{t.r.})^{\rm T} = \tilde\Psi^\dagger
\ee
or
\be\label{ur2}
\tilde\Psi^{t.r.} = \tilde\Psi^*\Sigma^{\rm T}
\:.\ee

These two last relations express how time reversal symmetry acts on the
scattering states. Let us now discuss this point in more detail.
If we consider a close graph ${\cal G}$ characterized by a matrix
$M(\gamma)$, the graph ${\cal G}^{t.r.}$ obtained by reversing the
magnetic fluxes is described by the matrix $M(\gamma)^{\rm T}$
[see eq. (\ref{MJean},\ref{RES3})].
If the original graph ${\cal G}$ has a spectrum $\{E_0,E_1,\cdots\}$,
the time reversed graph ${\cal G}^{t.r.}$ has the same set of energies~:
\be\label{hop}
E_n(\{\theta_{\mu\nu}\}) = E_n(\{-\theta_{\mu\nu}\})
\:.\ee
This very general feature is easily demonstrated for graphs since the
spectrum is given by $\det M(-E_n)=0$.
Note that stricktly speaking the energies are not in general even functions
of the magnetic field. It is only the full spectrum which is invariant under
time reversal symmetry, however we do not change the labels on both sides
of equation (\ref{hop}) to simplify the discussion.

If $\varphi_n(x)$ is an eigenstate of ${\cal G}$, then the basic fact that
the corresponding eigenstate of ${\cal G}^{t.r.}$ is its complex
conjugate,
\be
\varphi^{t.r.}_n(x) = \varphi_n(x)^*
\:,\ee
can be also easily recovered.
Indeed, if we call $\varphi_n$ the $V$-column vector gathering the values
of the wave function at the nodes,
$\varphi_n^{\rm T}=(\varphi_{n,1},\cdots,\varphi_{n,V})$,
it is a solution of $M(-E_n)\varphi_n=0$. Since $M(-E_n)^\dagger=-M(-E_n)$
we see that the eigenvector of $M(-E_n)^{\rm T}$ is $\varphi_n^*$.

Now we examine the connected graph and its stationary scattering states.
What is the relation between the scattering states of ${\cal G}$ and
${\cal G}^{t.r.}$~? Equation (\ref{ur1}) relates the values of
$\tilde\psi$ and $\tilde\psi^{t.r.}$ at all the vertices. Since we can
always
introduce an additional vertex on a bond, it can be written as~:
\be
\sum_\alpha \Sigma_\ab^* \: \tilde\psi^{t.r.\,(\alpha)}(x)
=\tilde\psi^{(\beta)}(x)^*
\:.\ee
If the graph is weakly coupled, the scattering matrix has a resonant
structure~:
\be
\Sigma_{\ab} \APPROX{E\sim E_n} -\delta_{\ab} +
\frac{2\I\sqrt{E_n}\, w_\alpha \varphi_n(\alpha)\,w_\beta
\varphi_n(\beta)^*}
     {E-E_n+\I\Gamma_n}
\:,\ee
where  $\Gamma_n=\sum_\alpha\Gamma_{n,\alpha}$ with
$\Gamma_{n,\alpha}=\sqrt{E_n}w_\alpha^2|\varphi_n(\alpha)|^2$ \cite{TexDeg02}.
The positions of the resonances and their widths remain unchanged
by reversing the fluxes. The scattering states behave like
\cite{TexDeg02}~:
\be
\tilde\psi^{(\alpha)}(x) \APPROX{E\sim E_n}
\frac{1}{\sqrt{\pi}}\,
\frac{\I E_n^{1/4} w_\alpha \varphi_n(\alpha)^*}{E-E_n+\I\Gamma_n}\,
\varphi_n(x)
\ee
near a resonance $E_n$, and the scattering states of ${\cal G}^{t.r.}$
behave like~:
\be
\tilde\psi^{t.r.\,(\alpha)}(x) \APPROX{E\sim E_n}
\frac{1}{\sqrt{\pi}}\,
\frac{\I E_n^{1/4} w_\alpha \varphi_n(\alpha)}{E-E_n+\I\Gamma_n}\,
\varphi_n(x)^*
\:.\ee
From the last equations we see that we must take care of the fact that
$\tilde\psi^{t.r.\,(\alpha)}(x)\neq\tilde\psi^{(\alpha)}(x)^*$.

\end{appendix}



\begin{thebibliography}{99}

\bibitem{ButImrLan83}
M.~B{\"u}ttiker, Y.~Imry, and R.~Landauer,
 Josephson behavior in small normal one-dimensional rings,
 Phys. Lett. A {\bf 96}, 365 (1983).

\bibitem{GefImrAzb84}
Y.~Gefen, Y.~Imry, and M.~Y. Azbel,
 Quantum oscillations and the Aharonov-Bohm effect for parallel
  resistors,
 Phys. Rev. Lett. {\bf 52}, 129 (1984).

\bibitem{ButImrAzb84}
M.~B{\"u}ttiker, Y.~Imry, and M.~Y. Azbel,
 Quantum oscillations in one-dimensional normal-metal rings,
 Phys. Rev. A {\bf 30}, 1982 (1984).

\bibitem{But85}
M.~B{\"u}ttiker,
 Small normal-metal loop coupled to an electron reservoir,
 Phys. Rev. B {\bf 32}, 1846 (1985).

\bibitem{WasWeb86}
S.~Washburn and R.~A. Webb,
 Aharonov-Bohm effect in normal metal. Quantum coherence and
  transport,
 Adv. Phys. {\bf 35}, 375 (1986).

\bibitem{VidMonDou00}
J.~Vidal, G.~Montambaux, and B.~Dou\c{c}ot,
 Transmission through quantum networks,
 Phys. Rev. B {\bf 62}, R16294 (2000).

\bibitem{PanAbiSerFouButVid01}
B.~Pannetier, C.~Abilio, E.~Serret, T.~Fournier, P.~Butaud, and J.~Vidal,
 Magnetic field induced localization in a two-dimensional
  superconducting wire network,
 Physica C {\bf 352} (2001).

\bibitem{NauFaiMaiEti01}
C.~Naud, G.~Faini, D.~Mailly, and B.~Etienne,
 Aharonov-Bohm cages in 2D normal metal networks,
 Phys. Rev. Lett. {\bf 86}, 5104 (2001).

\bibitem{Nau01}
C.~Naud,
 {\em Transport quantique dans les nanostructures},
 PhD thesis, Universit\'e Paris VI, 2001.

\bibitem{Fri52}
J.~Friedel,
 The distribution of electrons round impurities in monovalent metals,
 Philos. Mag. {\bf 43}, 153 (1952).

\bibitem{Kre53}
M.~G. Krein,
 Trace formulas in perturbation theory,
 Matem. Sbornik {\bf 33}, 597 (1953).

\bibitem{BetUhl37}
E.~Beth and G.~E. Uhlenbeck,
 Physica {\bf 4}, 915 (1937).

\bibitem{Hua63}
K.~Huang,
 {\em Statistical mechanics},
 John Wiley \& Sons, New York, 1963.

\bibitem{LanLif66e}
L.~D. Landau and E.~Lifchitz,
 {\em Physique statistique},
 Mir, 1966,
 Tome V.

\bibitem{But93}
M.~B{\"u}ttiker,
 Capacitance, admittance, and rectification properties of small
  conductors,
 J. Phys. Cond. Matter {\bf 5}, 9361 (1993).

\bibitem{GasChrBut96}
V.~Gasparian, T.~Christen, and M.~B{\"u}ttiker,
 Partial densities of states, scattering matrices and Green's
  functions,
 Phys. Rev. A {\bf 54}, 4022 (1996).

\bibitem{GraBut97}
T.~Gramespacher and M.~B{\"u}ttiker,
 Nanoscopic tunneling contacts on mesoscopic multiprobe conductors,
 Phys. Rev. B {\bf 56}, 13026 (1997).

\bibitem{SouSuz02}
S.~Souma and A.~Suzuki,
 Local density of states and scattering matrix in
  quasi-one-dimensional systems,
 Phys. Rev. B {\bf 65}, 115307 (2002).

\bibitem{SchTitBroBee02}
H.~Schomerus, M.~Titov, P.~W. Brouwer, and C.~W.~J. Beenakker,
 Microscopic versus mesoscopic local density of states in
  one-dimensional localization,
 Phys. Rev. B {\bf 65}, 121101 (2002).

\bibitem{GerPav88}
N.~I. Gerasimenko and B.~S. Pavlov,
 Scattering problems on noncompact graphs,
 Theor. Math. Phys. {\bf 74}, 230 (1988).

\bibitem{AvrSad91}
J.~E. Avron and L.~Sadun,
 Adiabatic quantum transport in networks with macroscopic components,
 Ann. Phys. (N.Y.) {\bf 206}, 440 (1991).

\bibitem{Ada92}
V.~Adamyan,
 Scattering matrices for microschemes,
 Oper. Theory: Adv. \& Appl. {\bf 59}, 1 (1992).

\bibitem{KosSch99}
V.~Kostrykin and R.~Schrader,
 Kirchhoff's rule for quantum wires,
 J. Phys. A: Math. Gen. {\bf 32}, 595 (1999).

\bibitem{KotSmi00}
T.~Kottos and U.~Smilansky,
 Chaotic Scattering on Graphs,
 Phys. Rev. Lett. {\bf 85}, 968 (2000).

\bibitem{TexMon01}
C.~Texier and G.~Montambaux,
 Scattering theory on graphs,
 J. Phys. A: Math. Gen. {\bf 34}, 10307 (2001).

\bibitem{Tex02}
C.~Texier,
 Scattering theory on graphs (2): the Friedel sum rule,
 J. Phys. A: Math. Gen. {\bf 35}, 3389 (2002).

\bibitem{Rot84}
J.-P. Roth,
 Une g\'en\'eralisation de la formule de Poisson,
 Publications math\'ematiques n$^o$23, Universit\'e de Haute Alsace,
  France, 1984.

\bibitem{Rot83}
J.-P. Roth,
 Spectre du Laplacien sur un graphe,
 C. R. Acad. Sc. Paris {\bf 296}, 793 (1983).

\bibitem{AkkComDesMonTex00}
E.~Akkermans, A.~Comtet, J.~Desbois, G.~Montambaux, and C.~Texier,
 On the spectral determinant of quantum graphs,
 Ann. Phys. (N.Y.) {\bf 284}, 10 (2000).

\bibitem{Des00}
J.~Desbois,
 Spectral determinant of Schr\"odinger operators on graphs,
 J. Phys. A: Math. Gen. {\bf 33}, L63 (2000).

\bibitem{Des01}
J.~Desbois,
 Spectral determinant on graphs with generalized boundary conditions,
 Eur. Phys. J. B {\bf 24}, 261 (2001).

\bibitem{Avr95}
J.~E. Avron,
 Adiabatic quantum transport,
 in {\em Quantum Fluctuations}, edited by E.~Akkermans, G.~Montambaux,
  J.-L. Pichard, and J.~Zinn-Justin, Proceedings of the Les Houches
  Summer School, Session LXI, (1995), Elsevier, Amsterdam, page 741.

\bibitem{But00}
M.~B{\"u}ttiker,
 Charge fluctuations and dephasing in Coulomb coupled conductors,
 in {\em Quantum mesoscopic phenomena and mesoscopic devices}, edited
  by I.~O. Kulik and R.~Ellialtioglu, Kluwer Academic
  Publisher, Dordrecht, (2000), volume 559, page 211,
 (cond-mat/9911188).

\bibitem{TexDeg02}
C.~Texier and P.~Degiovanni,
 Charge and current distribution in graphs,
 in preparation  (2002).

\bibitem{But02}
M.~B{\"u}ttiker,
 Charge densities and charge noise in mesoscopic conductors,
 Pramana J. Phys. {\bf 58}, 241 (2002),
 (cond-mat/0112330).

\bibitem{But96}
M.~B{\"u}ttiker,
 Dynamic conductance and quantum noise in mesoscopic conductors,
 J. Math. Phys. {\bf 37}, 4793 (1996).

\bibitem{PilBut02}
S.~Pilgram and M.~B{\"u}ttiker,
 Efficiency of mesoscopic detectors,
 Phys. Rev. Lett. {\bf 89}, 200401 (2002).

\bibitem{CleGirSto02}
A.~A. Clerk, S.~M. Girvin, and A.~D. Stone,
 Quantum-limited measurement and information in mesoscopic detectors,
 preprint cond--mat/0211001 (2002).

\bibitem{Bro98}
P.~W. Brouwer,
 Scattering approach to parametric pumping,
 Phys. Rev. B {\bf 58}, 10135 (1998).

\bibitem{AvrElgGraSad01}
J.~E. Avron, A.~Elgart, G.~M. Graf, and L.~Sadun,
 Optimal quantum pumps,
 Phys. Rev. Lett. {\bf 87}, 236601 (2001).

\bibitem{MosBut02}
M.~Moskalets and M.~B{\"u}ttiker,
 Dissipation and noise in adiabatic quantum pumps,
 Phys. Rev. B {\bf 66}, 035306 (2002).

\end{thebibliography}

\end{document}